\documentclass[12pt]{article}
\usepackage{cite}

\ifx\pdftexversion\undefined
  \usepackage[dvips]{graphics}
\else
  \usepackage[pdftex]{graphics}
\fi

\makeatletter
\newcommand{\eqref}[1]{(\ref{#1})}
\newcommand{\Chref}[1]{\expandafter\MakeUppercase\chaptername~\ref{#1}}
\newcommand{\Secref}[1]{\expandafter\MakeUppercase\secrefname~\ref{#1}}
\newcommand{\Appref}[1]{\expandafter\MakeUppercase\appendixname~\ref{#1}}
\newcommand{\Figref}[1]{\expandafter\MakeUppercase\figurename~\ref{#1}}
\newcommand{\Tbref}[1]{\expandafter\MakeUppercase\tablename~\ref{#1}}
    \let\secref\Secref  \let\appref\Appref
\let\figref\Figref  \let\tbref\Tbref
\newcommand{\secrefname}{Section}
\DeclareSymbolFont{AMSb}{U}{msb}{m}{n}
\DeclareSymbolFontAlphabet{\mathbb}{AMSb}
\newcommand{\inv}{^{\raise.15ex\hbox{${\scriptscriptstyle -}$}\kern-.05em 1}} 
\newcommand{\one}{\hbox{1\kern-.25em l}}                  
\newcommand{\CN}{\mathcal{N}}
\newcommand{\set}[1]{\mathbb{#1}}                         
\newcommand{\R}{\set{R}}
\newcommand{\Z}{\set{Z}}
\renewcommand{\H}{\set{H}}
\newcommand{\oper}[2][\mathrm]{\mathop{\kern\z@#1{#2}}\nolimits}
\newcommand{\group}[1]{\mathop{\kern\z@\mathrm{#1}}\nolimits}     
\newcommand{\U}{\group{U}}                                
\newcommand{\SU}{\group{SU}}      
\newcommand{\SO}{\group{SO}}
\newcommand{\SL}{\group{SL}}
\newcommand{\lie}[1]{\mathop{\kern\z@\mathit{#1}}\nolimits}       
\renewcommand{\u}{\lie{u}}                                
\newcommand{\su}{\lie{su}}        

\newcommand{\lsl}{\lie{sl}}
\newcommand{\opname}[1]{\mathop{\kern\z@\mathrm{#1}}\nolimits}    
\newcommand{\arctanh}{\opname{arctanh}}
\newcommand{\cosec}{\opname{cosec}}

\newcommand{\e}{\opname{e}}                               
\newcommand{\tr}{\opname{tr}}                             
\@addtoreset{equation}{section}
\renewcommand{\theequation}{\thesection.\arabic{equation}}
\renewcommand{\section}{\@startsection{section}{1}{\z@}%
                                    {-7ex \@plus -1ex \@minus -.2ex}%
                                    {2.5ex \@plus.2ex}%
                                    {\normalfont\large\scshape\centering}}          
\renewcommand{\subsection}{\@startsection{subsection}{2}{\z@}%
                                       {-5ex \@plus -1ex \@minus -.2ex}%
                                       {1.5ex \@plus.2ex}%
                                       {\normalfont\normalsize\scshape}}
\renewcommand{\subsubsection}{\@startsection{subsubsection}{3}{\z@}%
                                          {-4ex\@plus -1ex \@minus -.2ex}%
                                          {1.5ex \@plus .2ex}%
                                          {\normalfont\normalsize\scshape}}
\renewcommand{\paragraph}{\@startsection{paragraph}{4}{\z@}%
                                      {4ex \@plus1ex \@minus.2ex}%
                                      {-1em}%
                                      {\normalfont\normalsize\scshape}}
\renewcommand{\abstract}{\section*{\abstractname}}

\newcommand{\sectionname}{}                             

\renewcommand\@seccntformat[1]{\ignorespaces\csname #1name\endcsname\space
                               \csname the#1\endcsname.\quad}   
\renewcommand{\appendix}{\par
  \setcounter{section}{0}
  \setcounter{subsection}{0}%
  \renewcommand{\thesection}{\@Alph\c@section}
  \renewcommand{\sectionname}{\appendixname}}           
\def\eqnarray{%
   \stepcounter{equation}%
   \def\@currentlabel{\p@equation\theequation}%
   \global\@eqnswtrue
   \m@th
   \global\@eqcnt\z@
   \tabskip\@centering
   \let\\\@eqncr
   $$\everycr{}\halign to\displaywidth\bgroup
       \hskip\@centering$\displaystyle\tabskip\z@skip{##}$\@eqnsel
      &\global\@eqcnt\@ne$\;\hfil{##}$\hfil
      &\global\@eqcnt\tw@$\;\displaystyle{##}$\hfil\tabskip\@centering
      &\global\@eqcnt\thr@@ \hb@xt@\z@\bgroup\hss##\egroup
         \tabskip\z@skip
      \cr
}
\newdimen\captionmargin 
\newdimen\captionindent 
\newdimen\captionwidth 
\newcommand{\captionfont}{\slshape}
\newcommand\@captionlabel[1]{\textsc{#1:}\space}
\long\def\@makecaption#1#2{%
  \vskip\abovecaptionskip  
  \captionwidth\hsize 
  \advance\captionwidth -2\captionmargin
  \sbox\@tempboxa{\@captionlabel{#1}\captionfont #2}%
  \ifdim \wd\@tempboxa >\captionwidth
    \ifdim\captionindent>\z@ 
      \advance\captionwidth -\captionindent
      \hskip\captionindent
    \fi
    \hskip\captionmargin
    \parbox[t]{\captionwidth}{\leavevmode\hskip-\captionindent
      \@captionlabel{#1}\captionfont #2}%
  \else
    \global\@minipagefalse
    \hb@xt@\hsize{\hfil\box\@tempboxa\hfil}%
  \fi
  \vskip\belowcaptionskip}
\ifcase \@ptsize
\or 
\or 
\fi
  \divide\@tempdima\baselineskip
  \@tempcnta=\@tempdima
\makeatother

\begin{document}

%
%

\thispagestyle{empty}

\begin{flushright}
  \scshape\small
  WIS/35/03-DEC-DPP\\
  hep-th/0312209\\
  December 2003
\end{flushright}
\vskip15mm

\begin{center}

{\LARGE\scshape
Causal Structure of $d=5$ Vacua and Axisymmetric Spacetimes
\par}
\vskip15mm

\textsc{Bartomeu Fiol, Christiaan Hofman and Ernesto Lozano-Tellechea,}
\par\bigskip
{\itshape
  The Weizmann Institute of Science, Department of Particle Physics\\
  Herzl Street 2, 76100, Rehovot, Israel,}
\par\bigskip
\texttt{fiol, hofman, lozano@clever.weizmann.ac.il}

\end{center}
\vskip2cm

\section*{Abstract}

We study the structure of closed timelike curves (CTCs) for the near
horizon limit of the five dimensional BMPV black hole, in its
overrotating regime.  We argue that Bousso's holographic screens are
inside the chronologically safe region, extending a similar
observation of Boyda et al.~\cite{Boyda:2002ba} for G\"odel type
solutions. We then extend this result to quite generic axisymmetric
spacetimes with CTCs, showing that causal geodesics can't escape the
chronologically safe region. As a spin-off of our results, we fill a
gap in the identification of all maximally supersymmetric solutions of
minimal five dimensional supergravity, bringing this problem to a full
conclusion.

\newpage
\setcounter{page}{1}
%
%

\section{Introduction}
\label{sec:intro}

Much of the study of string theory solutions has been devoted to
static backgrounds, and many static compactifications preserving at
least a fraction of supersymmetry are relatively well understood. On
the other hand, the status of time dependent solutions in string
theory is much less satisfactory. Recently this situation has started
to change, with increasingly more attention being paid to
time-dependent backgrounds. Some of the approaches explored are
time-dependent orbifolds of Minkowski space~\cite{Simon:2002ma,
  Liu:2002ft, Cornalba:2002fi, Biswas:2003ku}, WZW
models~\cite{Elitzur:2002rt, Craps:2002ii}, double analytical
continuations of solutions~\cite{Aharony:2002cx} and
others~\cite{Buchel:2002kj}.  Together with the promise of bringing
string theory one step closer to the observed world (after all, the
Universe \emph{si muove}), time dependent backgrounds can present new
seemingly pathological features, absent from static compactifications.
For instance, in time-dependent orbifolds, new kinds of singularities
appear, and both time-dependent orbifolds of Minkowski space and WZW
models can have closed null curves~\cite{Simon:2002ma, Liu:2002ft} or
closed timelike curves (CTCs)~\cite{Khoury:2001bz, Elitzur:2002rt,
  Cornalba:2002fi}.

At this stage, the rules to deal with these pathologies are not clear,
but one might hope that by learning how string theory copes with them,
we can gain a better understanding of the theory, beyond the realm of
static solutions. In particular, the possible existence of closed
timelike curves in string theory raises many interesting conceptual
issues: does string theory actually allow solutions with closed
timelike curves, or do they have to be discarded? Is the chronology
protection conjecture~\cite{Hawking:pk} realized in string theory? If
so, what is the mechanism behind it? Typically, for time-dependent
backgrounds we currently lack the tools to address these questions.

A possible route to make progress is to consider stationary non-static
solutions in supergravity theories, which occupy an intermediate
position between static and time-dependent backgrounds. There are
plenty such solutions that are supersymmetric and everywhere regular,
and yet present CTCs\footnote{Conditions for the presence of CTCs were
  recently discussed in~\cite{Maoz:2003yv}.}. Among those, the
solutions of five dimensional supergravity (and their uplifts to ten
dimensions), have received special attention. This is in part due to
the characterization of all supersymmetric solutions of minimal five
dimensional supergravity~\cite{Gauntlett:2002nw}. Among the maximally
supersymmetric solutions, one finds a 5d generalization of the G\"odel
solution and the near horizon limit of the rotating
BMPV~\cite{Breckenridge:1996is} black hole (NH-BMPV). Both of those
solutions present closed timelike curves, at least in a range of
parameters.

The G\"odel type solution and its ten dimensional cousins have been
studied in considerable detail \cite{Boyda:2002ba, Harmark:2003ud,
  Drukker:2003sc, Hikida:2003yd, Brecher:2003rv, Brace:2003st,
  Brace:2003ww, Takayanagi:2003ps, Brecher:2003wq, Drukker:2003mg,
  Israel:2003cx}. Of particular relevance for the present work, is the
observation~\cite {Boyda:2002ba} that for G\"odel type solutions,
although there are closed timelike curves passing through every point
in spacetime, if we apply Bousso's prescription for holographic
screens~\cite{Bousso:1999cb} to these backgrounds, the induced metric
on the resulting screen is free of closed timelike curves. This gives
place to the speculation that there might be a holographic description
despite the presence of closed timelike curves on the classical
background.

In this work, we consider the other maximally supersymmetric solution
of $d=5$ minimal sugra with CTCs, the near horizon limit of the
overrotating BMPV black hole. The CTCs of the full overrotating BMPV
solution were studied in~\cite{Gibbons:1999uv}. We show that the
pattern of CTCs is very similar to that of the G\"odel solution, and
again Bousso's holographic screen is inside the chronologically safe
region. As a welcome bonus from our study, we bring to a full
conclusion the identification of all maximally supersymmetric
solutions of 5d minimal supergravity, filling a gap in the discussion
of~\cite{Gauntlett:2002nw} (see also~\cite{Chamseddine:2003yy}).
Namely, we explicitly show that the three solutions that were left
unidentified in~\cite{Gauntlett:2002nw} correspond to the near horizon
limit of the BMPV solutions, in the underrotating, critical and
overrotating cases, respectively.

After studying this five dimensional example in quite some detail, we
further consider more general metrics in arbitrary dimensions, where
closed timelike curves appear because of overrotation in different
planes, and show quite generally that causal (timelike or null)
geodesics cannot escape the chronologically safe region surrounding a
given observer. As a corollary, the corresponding holographic screens
are inside the chronologically safe regions, and the induced metrics
have no CTCs.

In this work we are concerned mostly with properties of classical
metrics with CTCs. It is worth noting that for some of the ten
dimensional sugra solutions, there are probe
computations~\cite{Emparan:2001ux, Dyson:2003zn, Drukker:2003sc}
suggesting that these backgrounds actually can't be built in string
theory.

The structure of the paper is as follows. In \secref{sec:5dvacua} we
review the status of vacua of five dimensional supergravity. In
\secref{sec:ctcnhbmpv} we study in detail the NH-BMPV solution. After
defining the notions we are going to use, like \emph{optical horizon}
or \emph{chronologically safe region}, we describe in detail the
structure of closed timelike curves in the NH-BMPV background, and
show that the optical horizon coincides with the boundary of the
chronologically safe region. In \secref{sec:ctcgeneral} we extend the
study of optical horizons and chronologically safe regions to quite
generic static and axisymmetric spacetimes, and we find that the
optical horizon never extends beyond the chronologically safe region;
furthermore, it generically coincides with the boundary of said
region, although we show examples where this is not the case. As a
corollary, Bousso's holographic screen for all these spacetimes is
inside the chronologicallty safe region, and the induced metric on
this screen has no CTCs.  Finally, we state our conclusions in
\secref{sec:concl}.

\section{NH-BMPV and the Complete identification of $d=5$ Vacua}
\label{sec:5dvacua}

One of the main purposes of this paper is the investigation of CTCs in
the background of the near horizon limit of the five dimensional
extreme rotating black hole (for short, we will henceforth use the
acronyms ``BMPV black hole'' for this black hole
spacetime~\cite{Breckenridge:1996is} and ``NH-BMPV'' for its
near-horizon limit\footnote{A classical reference in higher
  dimensional spinning black holes is~\cite{Myers:un}. See
  also~\cite{Horowitz:1995tm} for earlier work in supersymmetric
  rotating black holes. The five dimensional black hole we will be
  concerned with and its near horizon limit have been studied
  in~\cite{Chamseddine:1996pi,Kallosh:1996vy,Gauntlett:1998fz}.}).
This near horizon spacetime is
homogeneous~\cite{Alonso-Alberca:2002wr} and maximally supersymmetric
and, when the ``rotational'' parameter $j$ (to be defined below)
exceeds a certain critical value, it develops naked CTCs. We will
start with a detailed study of this spacetime itself. This, in turn,
will provide us with the answer to the problem of the identification
of all maximally supersymmetric solutions of minimal five dimensional
supergravity, an issue on which we now elaborate.

All bosonic solutions of pure $\CN=2$ (eight supercharges) five
dimensional supergravity preserving some fraction of supersymmetry
were characterized in~\cite{Gauntlett:2002nw}. In particular, all
\emph{maximally} supersymmetric solutions (which henceforth we will
call ``vacua''\footnote{Let us point out that there is no rigorous
  reason to call ``vacua'' these maximally supersymmetric solutions.
  These spacetimes enjoy a lot of (super)symmetries and, since all
  known maximally supersymmetric solutions are homogeneous, they have
  no ``core'' or ``energy locus'' in any physical sense. But this is
  only an analogy, and only in this sense one could call them
  ``vacua'' of supergravity theories.  However, since this shorthand
  is often used in the literature, here we will keep it for them.})
were obtained, and it was found that these include the following
spacetimes:
\begin{itemize}
  \item $AdS_2\times S^3$, which also arises as the near horizon limit
  of the non-rotating five dimensional extreme black
  hole~\cite{Chamseddine:1996pi}.

  \item $AdS_3\times S^2$, which is the near horizon limit of the
  extreme string solution~\cite{Gibbons:1994vm}. 

  \item The spacetime arising as the near horizon limit of the
  rotating BMPV black hole. 
  
  \item The homogeneous plane wave found in~\cite{Meessen:2001vx}. 

  \item A five dimensional analogue of the G\"odel universe.
\end{itemize}
In addition to those, three more solutions satisfying the necessary
conditions for maximal supersymmetry were also found
in~\cite{Gauntlett:2002nw}, but they were not
identified\footnote{I.e., those solutions were not shown to coincide
  (under, e.g., some coordinate transformation) with one of the
  spacetimes in the list above, nor neither shown to be genuinely new
  --- thus leaving, in this sense, an open question.}  neither
explicitly shown to be maxi\-mally supersymmetric (see Sections~5.3
and~5.4 of~\cite{Gauntlett:2002nw}). Subsequently, those unidentified
solutions where retrieved and shown to be indeed maximally
supersymmetric in~\cite{Chamseddine:2003yy}, where a complete
classification of all five dimensional vacua was obtained by
dimensional reduction from six dimensions, but, again, no explicit
identification of them was provided.  In~\cite{Gauntlett:2002nw}
and~\cite{Chamseddine:2003yy} the possibility that they might belong
to the NH-BMPV family was already anticipated, and below we are going
to show that this is indeed the case. In fact, we shall see that these
solutions correspond to the NH-BMPV spacetime in the
``underrotating'', ``critical'' (having closed lightlike curves) and
``overrotating'' (having CTCs) cases, resepctively\footnote{These
  NH-BMPV spacetimes are homogeneous and, as such, ``asymptote to
  themselves''.  This means that any attempt to define geometrical,
  conserved charges (as mass or angular momentum) with respect to
  their aymptotics~\cite{Abbott:1981ff} would yield, by construction,
  zero. However, we will often talk about ``rotating'' when referring
  to them, since the BMPV \emph{black hole} is of course rotating.  In
  the near horizon limit, the parameter $j$ we will be using below
  could be said to set the ``vorticity'' of spacetime, but not its
  angular momentum.}.

Among the five dimensional spacetimes in the list above, two of them
exhibit CTCs. These are the G\"odel universe and, generically, the
NH-BMPV spacetime. It was shown in~\cite{Lozano-Tellechea:2002pn} that
most maximally supersymmetric vacua of supergravity theories in
dimensions $d=4,5$ and 6 with eight supercharges can be related by
uplift and dimensional reduction. The fact that this has to be so for
the $d=5\leftrightarrow d=6$ vacua was explained
in~\cite{Chamseddine:2003yy}. What we are going to see below is that
the emergence of CTCs in five dimensions can be understood from the
identifications one makes in the six dimensional
vacua~\cite{Chamseddine:2003yy,Gutowski:2003rg}. 
This approach is what, following~\cite{Chamseddine:2003yy}, will in
the end provide us with an answer to the question about the complete
identification of maximally supersymmetric solutions in five
dimensions. In particular, we will focus on the near horizon BMPV
family. CTCs in the G\"odel case have already been
studied in the literature from several perspectives \cite{Herdeiro:2002ft,
Harmark:2003ud, Drukker:2003sc,Hikida:2003yd,Brecher:2003rv,Brace:2003st,
Brace:2003ww, Takayanagi:2003ps,Drukker:2003mg} and, concerning its lift 
to six dimensions, it was already
pointed out in~\cite{Chamseddine:2003yy} that it yields the maximally
supersymmetric six dimensional homogeneous wave
of~\cite{Meessen:2001vx}. As we will see below, the 6-dimensional
lifts of the whole near horizon BMPV family yield $AdS_3\times S^3$,
but with different global identifications dictated by the causal
structure of the five dimensional cases.

Next we describe the NH-BMPV family and its lift to six dimensions.
We will organize the possible cases according to the causal properties
of the five dimensional solutions. From the six dimensional
perspective they provide different Hopf fibrations of the $AdS_3$
factor, which we will call, respectively, ``spacelike'', ``timelike''
and ``lightlike''. We refer the interested reader to
\appref{app:hopf} for a more detailed discussion on those.

\subsection{The Near Horizon BMPV Family}

The near horizon BMPV black hole solution can be presented as:
\begin{equation}
  \label{nhbh}
  \left\{
  \begin{array}{r@{\;}c@{\;}l}
  ds^2 & = & \displaystyle{
             -\biggl(\frac{r}{R}\; dt - 
                    Rj(d\chi + \cos\theta\; d\varphi)\biggr)^2
             +\frac{R^2}{r^2}\; dr^2 + 4R^2 d\Omega_{(3)}^2}\, , \\[2.5ex]
  F    & = & \displaystyle{
                   -\frac{\sqrt{3}}{R}\; dt\wedge dr 
                   +\sqrt{3}Rj\sin\theta\; d\theta \wedge d\varphi
                   }\, , 
  \end{array}
  \right. 
\end{equation}
where 
\begin{displaymath}
  d\Omega_{(3)}^2= \frac{1}{4}\Bigl( d\Omega_{(2)}^2 
                   + (d\chi + \cos\theta\; d\varphi)^2\Bigr) = 
                   \frac{1}{4}\Bigl( d\theta^2+\sin^2\theta\; d\varphi^2 
                   + (d\chi + \cos\theta\; d\varphi)^2\Bigr)
\end{displaymath}
gives the usual Hopf-fibered form of the unit $S^3$ (i.e., the angles
$\theta$, $\varphi$ and $\chi$ take values in $[0,\pi]$,
$[0,2\pi]$ and $[0,4\pi]$, respectively). This solution (up to a
reparametrization $r\rightarrow r^2$ of the radial coordinate and with
a different choice for the parameters) is the near horizon limit of
the five dimensional supersymmetric extreme rotating black hole as
given in~\cite{Gauntlett:1998fz}\footnote{Our normalization of the
gauge field also differs from that of~\cite{Gauntlett:1998fz} by a
factor of two.}. The angular momentum of the black hole solution is
set by the dimensionless parameter $j$. For $j>1$,~\eqref{nhbh} has
closed timelike curves\footnote{This easy to check from the form of
the metric above. We will discuss in more generality the appearance of
CTCs in this spacetime and in a certain general class of metrics in
any spacetime dimension in Sections~\ref{sec:ctcnhbmpv}
and~\ref{sec:ctcgeneral}.}, while for $j=1$ there are closed lightlike
curves. We will now look at the six dimensional lift of these solutions
in the different ranges of $j$.  For $j<1$ we will write $j=\sin\xi$,
while for $j>1$ we will parametrize it as $j=\cosh\xi$.

\subsection{Six Dimensional Lifts of NH-BMPV}

Let us first record here the lifting rules to six dimensions. For a
general five dimensional solution given by a line element $ds^2$ and 
two-form $F=dA$, the uplifting rules to the minimal six dimensional
supergravity theory (whose bosonic field content is given by the
six dimensional metric and an anti-self-dual three form $H=dB$) are
given by~\cite{Lozano-Tellechea:2002pn}:
\begin{equation}
  \label{LiftTo6}
  \left\{
  \begin{array}{r@{\;}c@{\;}l}
  ds_{(6)}^2 & = & \displaystyle ds^2 + 
                 \biggl(dw + \frac{1}{\sqrt{3}}A\biggr)^2\, , \\[2ex]
  B_{\mu w}  
             & = & \displaystyle \frac{1}{\sqrt{3}}A_{\mu}\, , 
  \end{array}
  \right.
\end{equation}
where $w$ is the ``sixth coordinate''. The remaining components of the
two-form potential can be calculated from the anti-self-duality
constraint. All five and six dimensional solutions that we will
describe have, respectively, nontrivial two- and three-form fields.
We will omit from now on the expressions for the gauge fields,
since our main interest concerns the causal structure and the global
identifications of the spacetime metrics.

\subsubsection{Six Dimensional Lift for $j<1$}

We first look at the case $j<1$, which has no causal singularities.
After writing $j=\sin\xi$ and performing the rescaling
\begin{equation}
  \label{tRescalingUR}
  t \rightarrow \cos\xi\; t  
\end{equation}
we can write the five dimensional metric as\footnote{For our purposes,
it is more convenient to take the period of the $\chi$-angle entering
in the one-form $\omega^3\equiv d\chi+\cos\theta\; d\varphi$ fixed to
$4\pi$.}:
\begin{equation}
  \label{nhbhspace}
  ds^2 = -\frac{r^2}{R^2}\; dt^2 + \frac{R^2}{r^2}\; dr^2 
         + R^2d\Omega_{(2)}^2 
         + R^2\cos^2\xi\biggl(d\chi + \cos\theta\; d\varphi
         + \tan\xi \frac{r}{R^2}\; dt\biggr)^2\, ,  
\end{equation}
making clear that this spacetime is a $\U(1)$ fibration over
$AdS_2\times S^2$. In fact, it was shown
in~\cite{Alonso-Alberca:2002wr} that the $j<1$ NH-BMPV spacetime is
homogeneous and, \emph{locally},
\begin{displaymath}
  \mbox{NH-BMPV}\simeq\frac{\SO(2,1)\times \SO(3)}{\U(1)_\xi}\, ,
\end{displaymath}
with $\xi$ determining the relative weight of the $\U(1)$ action on
each factor. This already indicates that we are dealing with a
quotient of $AdS_3\times S^3$ and explains why the $j<1$ subfamily
admits a local metric which interpolates between those of $AdS_2\times
S^3$ at $\xi=0$ and $AdS_3\times S^2$ at $\xi=\pi/2$ (this is not
apparent in these coordinates, though --- see the discussion
in~\cite{Lozano-Tellechea:2002pn,Alonso-Alberca:2002wr}). This
interpolation has also been pointed out
in~\cite{Gauntlett:2002nw,Chamseddine:2003yy}. One of the things that
we will clarify later on is the fact that, although the $j=0$
case is indeed $AdS_2\times S^3$, the $j=1$ case is not $AdS_3\times
S^2$. Only the $j\to1$ \emph{limit} of the $j<1$ subfamily admits a
metric which is locally that of $AdS_3\times S^3$. In fact, we will
show that the $AdS_3\times S^2$ vacuum is not smoothly connected to
the NH-BMPV family. The fact that the $j=1$ limit of both the
underrotating and overrotating NH-BMPV subfamilies
(Eqns.~\eqref{nhbhspace} and~\eqref{nhbhtime}) is singular can be
interpreted as an indication of the peculiar properties of the $j=1$
case.

After these considerations about this five dimensional spacetime, let
us write its lift to six dimensions. Using the
KK-rules~\eqref{LiftTo6} and further defining the rotated angles
\begin{equation}
  \pmatrix{ u  \cr \psi } = 
  \pmatrix{ \sin\xi  & \cos\xi \cr 
            -\cos\xi & \sin\xi }
  \pmatrix{ \cos\xi\; \chi \cr  w/R }\, ,  
\end{equation}
we get the six dimensional metric 
\begin{equation}
  \label{6DimLiftPoincSpace}
  ds_{(6)}^2 = -\frac{r^2}{R^2}\; dt^2 + \frac{R^2}{r^2}\; dr^2  
               +R^2\biggl(du + \frac{r}{R^2}\; dt\biggr)^2 
               + 4R^2d\Omega_{(3)}^2\, , 
\end{equation}
where now $4d\Omega_{(3)}^2=d\Omega_{(2)}^2 + (d\psi - \cos\theta\;
d\varphi)^2$.  We verify (see \appref{app:poincare}) that this
spacetime is locally $AdS_3\times S^3$, and that the metric in this
form does no longer depend on the ``rotational'' parameter $j$, since
its six dimensional meaning is the choice of the KK compactification
direction. Using the coordinate transformation given
by~\eqref{trafo1} we can write this metric in global coordinates and
see that the $AdS_3$ factor precisely corresponds to the spacelike
Hopf fibration of $AdS_3$ over $AdS_2$ (along the coordinate that we
now call $\tilde{\psi}$) discussed in \appref{app:hopf}:
\begin{equation}
  \label{UR6DimLift}
  ds_{(6)}^2 = -R^2\cosh^2\frac{\rho}{R}\; d\tau^2 + d\rho^2 
               +R^2\biggl(d\tilde{\psi} 
               +\sinh\frac{\rho}{R}\; d\tau\biggr)^2\, 
               +4R^2d\Omega_{(3)}^2. 
\end{equation}
It is worth mentioning that, to make contact with five dimensions, we
have to keep the ``sixth coordinate'' compact. When keeping $w$
compact, its periodicity has to be chosen so that the period of $\psi$
is $4\pi$ in order to have a regular space. This automatically sets
the period of $\tilde{\psi}$ and ensures a regular space and a
well-defined $\U(1)$ action along $\tilde{\psi}$ whenever $j<1$. Note
that the Hopf direction $\tilde{\psi}$ is spacelike. It is crucial to
take this point of view on the compactness of the six dimensional
backgrounds in order to make contact with the analysis that we will
perform in \secref{ClassAndRedFrom6Dim} and, of course, it also
applies to the $j>1$ and $j=1$ cases.

\subsubsection{Six Dimensional Lift for $j>1$}

Next we look at the case $j>1$, which has CTCs. After setting 
$j=\cosh\xi$ and the rescaling
\begin{equation}
  \label{tRescalingOR}
  t \rightarrow \sinh\xi\; t\, , 
\end{equation}
one can write the metric as
\begin{equation}
  \label{nhbhtime}
  ds^2 = \frac{r^2}{R^2}\; dt^2 + \frac{R^2}{r^2}\; dr^2 
         + R^2d\Omega_{(2)}^2 
         - R^2\sinh^2\xi\biggl(d\chi + \cos\theta d\varphi
         - \coth\xi \frac{r}{R^2}\; dt\biggr)^2\, .    
\end{equation}
We can see that this spacetime is a $\U(1)$ fibration over the
hyperbolic plane $\H_2$ times the sphere $S^2$. The six dimensional
lift now reads:
\begin{equation}
  ds_{(6)}^2 = \frac{r^2}{R^2}dt^2 + \frac{R^2}{r^2}dr^2  
               -R^2\biggl(du + \frac{r}{R^2}dt\biggr)^2 
               + R^2d\Omega_{(3)}^2\, , 
\end{equation}
where $d\Omega_{(3)}^2$ is as in~\eqref{6DimLiftPoincSpace}, but now
with the ``boosted'' angles defined as:
\begin{equation}
  \pmatrix{ u \cr \psi } = 
  \pmatrix{ \cosh\xi  & -\sinh\xi \cr 
            -\sinh\xi &  \cosh\xi }
  \pmatrix{ \sinh\xi\; \chi \cr w/R }\, .  
\end{equation}
This spacetime is again $AdS_3\times S^3$ locally, but global
identifications are now completely different. By using the coordinate
transformation~\eqref{trafo2} we can write this in global coordinates
and find it to coincide with the timelike Hopf fibration of the $AdS_3$
factor:
\begin{equation}
  \label{OR6DimLift}
  ds_{(6)}^2 =  R^2\cosh^2\frac{\rho}{R}\; d\tau^2 + d\rho^2 
               -R^2\biggl(d\tilde{\psi} 
               +\sinh\frac{\rho}{R}\; d\tau\biggr)^2\, 
               +4R^2d\Omega_{(3)}^2
\end{equation}
(compare to~\eqref{UR6DimLift}). The $\xi$-dependence also goes away
in the six dimensional spacetime. Again, we can always choose the
appropriate period for $w$ in order to have a smooth space with a
well-defined $\U(1)$ action along the Hopf direction $\tilde{\psi}$,
which now is timelike.

\subsubsection{Six Dimensional Lift for $j=1$}

Note that~\eqref{nhbhspace} and~\eqref{nhbhtime} are, in principle,
not smoothly connected in parameter space, since the above
rescalings of the time coordinate are not allowed when $j=1$ (and, in
fact, both five dimensional metrics as written below are singular in
that limit). 
The lift of~\eqref{nhbh} when $j=1$ gives the metric:
\begin{equation}
  ds_{(6)}^2 = 2r(d\chi-d\psi)dt + \frac{R^2}{r^2}dr^2  
               + R^2d\Omega_{(3)}^2\, ,
\end{equation}
where now 
$4d\Omega_{(3)}^2=d\Omega_{(2)}^2 + (d\psi - \cos\theta d\varphi)^2$. 
To get this form we have just defined
\begin{equation}
  w \rightarrow R\psi\,  
\end{equation}
after the uplift. Putting $\tilde{\psi}\equiv\chi-\psi$ we make
contact with the $AdS_3$ metric~\eqref{poincare1} in the
case $\alpha=0$. If we further define
\begin{displaymath}
  \rho=R\; \log\frac{r}{R}\, ,
\end{displaymath}
we have the metric
\begin{equation}
  \label{Crit6DimLift}
  ds_{(6)}^2 = 2e^{\rho/R}dtd\tilde{\psi} + 
               d\rho^2 + 4R^2d\Omega_{(3)}^2\, , 
\end{equation}
which, from~\eqref{hopfnull}, we can again identify locally with
$AdS_3\times S^3$, but now with the $AdS_3$ factor given by the
lightlike Hopf fibration discussed in \appref{app:hopf}. We see
that, also in this case, we have a well-defined $\U(1)$ action along
the null coordinate $\tilde{\psi}$ with period equal to $4\pi$.

\subsection{Classification and Reduction from 6 Dimensions}
\label{ClassAndRedFrom6Dim}

We have seen that the near horizon limits of the five dimensional
extreme rotating black hole can be related to reductions of
$AdS_3\times S^3$ with various identifications, and so far we have
related their explicit local metrics in five and six
dimensions. However, the relevant piece of information in order to
analyze both the classification of five dimensional vacua and their
causal structure concerns \emph{global} issues. We now elaborate on
this, following almost in parallel the lines and notation established
in~\cite{Chamseddine:2003yy}. 

We have shown in the preceding section that the $j<1$, $j>1$ and $j=1$
cases of the NH-BMPV family, when lifted to six dimensions, 
can be seen locally as quotients 
\begin{equation}
  \label{AdS3S3OverU1}
  \frac{AdS_3\times S^3}{\U(1)}\, , 
\end{equation}
but with different choices for the direction of the $\U(1)$. It turns
out that this is really helpful in order to understand these five
dimensional vacua, since all possible maximally supersymmetric
solutions in $d=5$ were shown in~\cite{Chamseddine:2003yy} to be given
by all possible ``spacelike quotients'' of all possible six dimensional
vacua, such as $AdS_3\times S^3$. All six dimensional vacua are given
by Lorentzian Lie groups~\cite{Chamseddine:2003yy}, and by ``spacelike
quotients'' we mean those which are in one-to-one correspondence with
the different spacelike one-parameter subgroups of these six dimensional
vacua. In order to see this, what we are going to do next is, in a
sense, the opposite analysis of the one in the last Section. We are
going to ``dimensionally reduce'' $AdS_3\times S^3$, investigating the
properties of all possible spacelike quotients and paying special
attention to global aspects. To do so, we will identify $AdS_3$ with
the universal cover of $\SL(2)$, while $S^3$ will be seen as
$\SU(2)$. From the point of view of $AdS_3$ --- when seen (as usual) as
its universal covering --- all freely acting 1-parameter subgroups are
noncompact. So, strictly speaking, we have to look at quotients of the
kind:
\begin{equation}
  \frac{\widetilde{\SL}(2)\times\SU(2)}{\R}\, . 
\end{equation}
We will take the convention that $\R$ always acts by a right action of
$\widetilde{\SL}(2)\times\SU(2)$ on itself. The direction in the
second factor can always be chosen to be proportional to a fixed Lie
algebra generator ($\kappa$, say) of $\su(2)$.  However, in
$\widetilde{\SL}(2)$ we can choose inequivalent generators. Let us
denote, as in~\cite{Chamseddine:2003yy}, by $\sigma$, $\nu$ and $\tau$
a spacelike, lightlike and timelike generator of $\lsl(2)$
respectively; e.g. given by $\sigma=\hat{\tau}^1$,
$\tau=\hat{\tau}^2$, $\nu=\hat{\tau}^1+\hat{\tau}^2$ (see
\appref{app:hopf} for notation). Then the possible spacelike
generators are of the form (1) $a\sigma+b\kappa$, (2) $\nu+\kappa$ or
(3) $a\tau+b\kappa$, where $a$ and $b$ are arbitrary real weights,
although one must realize that in the last case we need $b^2>a^2>0$ in
order to have a spacelike subgroup. For the first case we will
consider separately the cases ($1'$) $\sigma$ and ($1''$) $\kappa$.

To see the $d=5$ solutions as genuine KK-reductions from $d=6$ we need
to describe them not as a quotient by $\R$, but rather as a quotient
by $\U(1)$.  Note that, except for the case ($1'$), the subgroup always
acts on $S^3$. As this space is compact, there is a $\Z$ subgroup that
acts trivially on this factor. Therefore this $\Z$ subgroup only acts
non trivially on $AdS_3$ and we can divide it out separately. That is,
we can write:
\begin{equation}
  \frac{AdS_3\times S^3}{\R} = 
  \biggl(\frac{AdS_3}{\Z}\times S^3\biggr)/\U(1)\, , 
\end{equation}
the $\U(1)$ here being the one acting along the ``sixth coordinate''
in the different lifts~\eqref{UR6DimLift}, \eqref{OR6DimLift}
and~\eqref{Crit6DimLift}, i.e.\ the $\U(1)$ in the local description
provided by~\eqref{AdS3S3OverU1}. The $\Z$ action is of course
generated by the same Killing vector as $\R$, but with a period
dictated by the one in $S^3$. For example, if $\kappa$ is normalized
such that it has period $2\pi$ on $S^3$, we find that the period of
$\Z$ on $AdS_3$ is $2\pi a/b$ in case (1), etc.  In case ($1'$) the
Killing vector acts only on $AdS_3$, and therefore we could have
chosen any $\Z$ subgroup to begin with.  In the case ($1''$) we are
forced from the start to have an action by $\U(1)$ instead of $\R$, as
it acts only on the compact space $S^3$. The different possibilities
and the relations to extremal black holes and $d=4$ Robinson-Bertotti
(to be discussed below) are summarized in \tbref{tb:red}.

\begin{table}[htbp]
  \begin{center}
  \begin{tabular}{lllll}
  case  & generator & $j$ range & NH of $d=5$ extremal BH 
  & $d=4$ reduction to RB \\\hline
  (1)   & $a\sigma+b\kappa$ & $0<j<1$ 
  & physical rotating & dyonic $j=\sin\xi$ \\
  ($1'$)  & $\sigma$ & $j=0$ 
  & non-rotating ($AdS_2\times S^3$)& electric \\
  ($1''$) & $\kappa$ &  
  & NH of black string ($AdS_3\times S^2$)& magnetic \\
  (2) & $\nu+\kappa$ 
  & $j=1$ & critical rotating & \\
  (3) & $a\tau+b\kappa$ & $j>1$ & overrotating & 
  \end{tabular}
  \caption{The possible reductions of $AdS_3\times S^3$ and their 
  relation to $d=5$ and $d=4$ maximally supersymmetric solutions.}
  \label{tb:red}
  \end{center}
\end{table}
The identification of these quotients with the supergravity solutions
belonging to the NH-BMPV family follows from the spacelike, timelike
or null properties of the identifications discussed here and those of
the six dimensional lifts discussed in the previous section. The
identification of case ($1''$) with $AdS_3\times S^2$ is obvious.

Let us now analyze how these different solutions are connected.  We
can continuously go from case (1) to case ($1'$). Indeed, thought of as
the action on a cylinder (the noncompact direction being that in
$AdS_3$, the circle being in $S^3$, and the one-parameter subgroup
being a line winding the cylinder), the only thing that happens when
$j\to0$ (or $b\to0$) is that the action has a longer and longer period
over the noncompact direction --- the endpoint is a decompactification
(which is very much like the limit of a circle to the real line).
However, we \emph{can not} continuously go from case (1) to case
($1''$), even though the Killing vector changes continuously. The reason
is that the quotient group goes discontinuously from $\R$ to $\U(1)$.
If we try to take the limit, the extra factor of $\Z$ acting on
$AdS_3$ will reduce the circle (the one which should arise after the
identification) to zero size, and make a degenerate limit. A similar
reasoning holds for the relation to case (3), taking $a\to0$: this is
not a well defined limit and therefore ($1''$) is, again, not
continuously connected to case (3).

When from e.g.\ case (1) we take $j\to1$, what we should do is to zoom
into the circle that is vanishing. The way to do this is as in Matrix
theory, by applying a large boost, which becomes infinite in the $j=1$
limit. Then the zero-size circle is boosted to a finite size (but zero
proper length) lightlike circle, and what we end up with is case
(2). Therefore we find that case (2) is continuously connected to case
(1) and, similarly, case (2) is also continuously connected to case
(3).  An indication of this fact is that~\eqref{nhbh} provides a single, 
never degenerate metric for the whole NH-BMPV family. 

Summarizing, we find that all solutions live in the same family,
except for case ($1''$). This is why, as already advanced, the
$AdS_3\times S^2$ vacuum cannot be continuously connected to the
NH-BMPV family. Note that this can be physically understood from the
BPS configurations that give rise to these near horizon spacetimes:
$AdS_3\times S^2$ is the only solution that is not the near-horizon
limit of a BH, but rather of the black string. 

Finally, let us mention that, from this six dimensional point of view,
the emergence of closed lightlike curves and CTCs in cases (2) and (3)
is clear: as discussed in~\cite{Chamseddine:2003yy} (see
also~\cite{Figueroa-O'Farrill:2002tb} for similar reductions), this is
due to the null and timelike character of the Killing vectors $\nu$
and $\tau$ generating the identifications.  Even if the ``sixth
circle'' is spacelike, it is enforcing identifications inside the
lightcone in the dimensionally reduced space.

\subsubsection{Relation to Four Dimensional Robinson-Bertotti Spacetimes}

We have seen how the NH-BMPV family is related to $AdS_3\times S^3$ in
six dimensions. In~\cite{Lozano-Tellechea:2002pn} it was shown that
five dimensional vacua are also related to four dimensional ones
since, in general, they can be dimensionally reduced to four
dimensions without breaking any supersymmetry, hence giving rise to
vacua of pure $\CN=2$, $d=4$ supergravity theory.  In particular, it
was shown that the $j<1$ NH-BMPV family and the $AdS_3\times S^2$
vacuum, when reduced to $d=4$, both give rise to near horizon limits
of Robinson-Bertotti spacetimes (which are the near horizon limit of
the Reissner-Nordstr\"om black holes and whose metrics are those of
$AdS_2\times S^2$). In $d=4$ we have a whole family of RB spacetimes
due to four dimensional electric-magnetic duality. In this way, the
$j=0$ case ($AdS_2\times S^3$) gives rise to the purely electric RB
solution, while the $j<1$ BH-BMPV gives rise to a general dyonic RB
spacetime (with ratio between electric and magnetic charges precisely
determined by $j=\sin\xi$, where now $\xi$ exactly coincides with the
angle parametrizing electric-magnetic duality ``rotations'' in four
dimensions\footnote{The classical four dimensional electric-magnetic
  duality group of the $\CN=2$ theory is $\SO(2)$}). On the other
hand, the lift of the \emph{purely magnetic} RB solution was shown
in~\cite{Lozano-Tellechea:2002pn} to correspond to $AdS_3\times S^2$.
These results are summarized in \tbref{tb:red}. It is instructive
to take a closer look at this, since we have just shown that the five
dimensional $AdS_3\times S^2$ vacuum is not continuously connected to
the $j<1$ NH-BMPV family, while in four dimensions electric-magnetic
duality of course interpolates smoothly between the purely electric
and purely magnetic cases.

In four dimensions the situation is as follows. It is a well-known
fact that electric-magnetic duality acts locally on the gauge
invariant field strength, but in a highly nonlocal way on the gauge
potential. Purely electric and purely magnetic RB spacetimes are
smoothly connected when written in terms of the field strength.
However, when written in terms of the gauge potential (or more
conveniently the dual gauge potential $\tilde{A}_{(4)}$), the
situation is more subtle.

When we approach the monopole configuration in the limit $j\to1$, the
gauge transformation for $\tilde{A}_{(4)}$ needed to remove the Dirac
string singularity --- which is a global issue --- has to be
periodically identified with a period which is vanishing in the limit.
If one wishes to couple the four dimensional theory to magnetic
charges, it is easily verified that what is happening is that the unit
of magnetic charge diverges in the limit. Only in the purely magnetic
case, when the dual potential $\tilde{A}_{(4)}$ is ``purely
electric'', everything is well defined.

From the five dimensional point of view we are forced to use 
$\tilde A_{(4)}$, as we can not write the five dimensional 
metric in terms of gauge invariant quantities. The metric 
in five dimensions is in fact given by 
\begin{displaymath}
  ds^2 \sim ds_{(4)}^2 + \left(dx^5 + \tilde{A}_{(4)}\right)^2\, ,
\end{displaymath}
and hence we see that the periodicity of the gauge field in four 
dimensions corresponds to the periodicity of the fifth coordinate $x^5$.

\subsubsection{Explicit Identification of Five Dimensional Solutions}

The considerations above show that all possible reductions of the
6-dimensional $AdS_3\times S^3$ vacuum are in the NH-BMPV family.
Since all possible inequivalent reductions of the other nontrivial
6-dimensional vacuum (namely, the homogeneous plane wave
of~\cite{Meessen:2001vx}) give the maximally supersymmetric
5-dimensional plane wave or the G\"odel
universe~\cite{Chamseddine:2003yy}, we conclude that the different
NH-BMPV cases complete all possible five dimensional vacua.  We have
thus identified them. However, for the sake of completeness, let us
explicitly show that the different cases of NH-BMPV can be related to
the three unidentified solutions of~\cite{Gauntlett:2002nw}. For
convenience, we now set to unit the scale parameter $R$.

\paragraph{$\bullet$ $j<1$:}
Starting from the solution~\eqref{nhbhspace} for $j<1$ and using
essentially the coordinate transformation~\eqref{trafo1}, followed by
$R=\sinh\rho$, we find the solution (5.118)
of~\cite{Gauntlett:2002nw},
\begin{equation}
 ds^2 = -(R^2+1){dt'}^2 + \frac{dR^2}{R^2+1} + \cos^2\xi(d\psi + \tan\xi\;
 Rdt'+ \cos\theta\; d\phi )^2 + d\theta^2 +\sin^2\theta\;d\phi^2\, .
\end{equation}

\paragraph{$\bullet$ $j>1$:}
Starting from the solution~\eqref{nhbhtime} for $j>1$ and using the
coordinate transformation~\eqref{trafo3}, followed by $R=\cosh\rho$,
we find the solution~(5.113) of~\cite{Gauntlett:2002nw},
\begin{equation}
 ds^2 = \frac{dR^2}{R^2-1} + (R^2-1){d\chi}^2 - \sinh^2\xi (d\psi -
 \coth\xi\; Rd\chi + \cos\theta\; d\phi )^2 + d\theta^2
 +\sin^2\theta\;d\phi^2\, .
\end{equation}

\paragraph{$\bullet$ $j=1$:}
Finally, starting from the $j=1$ solution as given by~\eqref{nhbh} and
making the coordinate transformation given by
\begin{equation}
r= 2r' \cos^2 t'\, , \qquad
t= \tan t'\, ,     \qquad
\chi= x + \frac{\tan t'}{r'}\, ,
\end{equation}
followed by a rescaling of $t'$, we find the solution~(5.102)
of~\cite{Gauntlett:2002nw}:
\begin{equation}
 ds^2 = - (1+{r'}^2){dt'}^2 + \frac{{dr'}^2}{{r'}^2} 
        + 2r'dt'(dx+\cos\theta\; d\phi) +d\theta^2+\sin^2\theta\;d\phi^2\, . 
\end{equation}

\section{CTCs and Optical Horizons in the Near Horizon BMPV}
\label{sec:ctcnhbmpv}

In this section we will consider the near horizon BMPV solution in the
overrotating case $j>1$. We noted in the preceding section that these
solutions have closed timelike curves. In the present section we will
study the domains with CTCs in detail, and determine the
chronologically safe region enclosing a given observer. We will then
determine the regions that can be reached by causal (timelike and
null) geodesics. We will show that for a given observer, no causal
geodesic going through his world-line escapes the chronologically safe
region. As a consequence, Bousso's holographic screen for a given
observer will be completely inside the chronologically safe region.
Thus the NH-BMPV solution provides another example where the
conjecture~\cite{Boyda:2002ba} that holography acts as the chronology
protection agent might apply.

\subsection{Metric and Isometries}

The metric of the NH-BMPV for $j>1$ was written in~\eqref{nhbhtime} in
Poincar\'e coordinates.  It will be useful to use~\eqref{trafo3},
together with some straightforward rescalings, to write this metric in
global coordinates in the following form
\begin{eqnarray}\label{ctcmetric}
  ds^2 &=& d\rho^2 + R^2\sinh^2\frac{\rho}{R}\;d\phi^2 
  + R^2(d\theta^2 + \sin^2\theta\;d\varphi^2) 
\nonumber\\
&&  -\Bigl(d\tau+2jR\sinh^2\frac{\rho}{2R}\;d\phi-2\sqrt{j^2-1}R\sin^2
\frac{\theta}{2}\;d\varphi\Bigr)^2. 
\end{eqnarray}
We renamed the coordinates $\psi$ to $\tau$, and $\tau$ to $\phi$, as
the former is always a timelike coordinate for $j>1$.  We also shifted
the coordinate $\tau$ and used $\cosh\rho-1=2\sinh^2\frac\rho2$ and
$\cos\theta-1=-2\sin^2\frac\theta2$. Note that after this shift the
coordinates $\phi$ and $\varphi$ are genuinely periodic with period
$2\pi$ (at fixed $\rho,\theta,\tau$), as they are angles in polar
coordinates with radii $\rho$ and $\theta$ respectively.  This
solution clearly exhibits a fibration of the $\tau$ line over the
product of a hyperbolic plane parametrized by polar coordinates
$(\rho,\phi)$ and a sphere parametrized by $(\theta,\varphi)$, of
respective curvatures $\mp 1/R$. This metric can be seen as combining
the G\"odel solutions of~\cite{Reboucas:hn} fibered over the
hyperbolic plane and the sphere in a single metric.  In this section
we will take $\tau$, being a global time coordinate, to be
noncompact. We will comment on this choice at the end of this
section.

The metric~\eqref{ctcmetric} is spacetime homogeneous and has a large
number of symmetries. The isometry group is
$\U(1)\times\SU(2)\times\widetilde{\SL}(2)$~\cite{Gauntlett:1998kc,Gauntlett:1998fz},
generated by the Killing vectors:
\begin{eqnarray}
\label{u1killing}
  \u(1):  && \quad\xi_0 = \partial_\tau\, , \\
\label{su2killing}
  \su(2): && \left\{
  \begin{array}{r@{\;}c@{\;}l}
    \xi_1 &=& \displaystyle 
              \sin\varphi\;\partial_\theta 
              + \cos\varphi\cot\theta\;\partial_\varphi 
              - \sqrt{j^2-1}R\cos\varphi\tan\frac\theta2\;\partial_\tau\, , 
      \\[1.5ex]
    \xi_2 &=& \displaystyle 
              \cos\varphi\;\partial_\theta 
              - \sin\varphi\cot\theta\;\partial_\varphi 
              + \sqrt{j^2-1}R\sin\varphi\tan\frac\theta2\;\partial_\tau\, , 
      \\[1.5ex]
    \xi_3 &=& \displaystyle 
              \partial_\varphi-\sqrt{j^2-1}R\,\partial_\tau\, ,
  \end{array}
  \right. \\
\label{sl2killing}
  \lsl(2):&& \left\{
  \begin{array}{r@{\;}c@{\;}l}
    \xi_4 &=& \displaystyle 
              R\sin\phi\;\partial_\rho 
              + \cos\phi\coth\frac{\rho}{R}\;\partial_\phi
              + jR\cos\phi\tanh\frac{\rho}{2R}\;\partial_\tau\, ,
      \\[1.5ex]
    \xi_5 &=& \displaystyle 
              R\cos\phi\;\partial_\rho 
              - \sin\phi\coth\frac{\rho}{R}\;\partial_\phi 
              - jR\sin\phi\tanh\frac{\rho}{2R}\;\partial_\tau\, ,
      \\[1.5ex]
    \xi_6 &=& \displaystyle 
              \partial_\phi-jR\,\partial_\tau\, .
  \end{array}
  \right.
\end{eqnarray}

As the metric~\eqref{ctcmetric} is spacetime homogeneous, all points
in spacetime are physically equivalent and, without any loss of
generality, we can consider a comoving observer located at
$(\rho,\theta)=(0,0)$.  The spacetime homogeneity of the metric will
also be very helpful in studying the occurrence of geodesics and CTCs.

\subsection{Definitions}

Before proceeding, we collect here some definitions that we will using
in the rest of the paper.

\emph{Holographic screens:} According to Bousso's
prescription~\cite{Bousso:1999cb}, given a spacetime, we choose a
foliation into null hypersurfaces. For each hypersurface, we consider
the expansion of null geodesics, and we mark the points in the
hypersurface where that expansion vanishes. If we do that for each
null hypersurface, the set of all marked points constitutes the
holographic screen. According to the conjecture
of~\cite{Bousso:1999cb}, the proper area of the screen (in Planck
units) at each slice gives the number of degrees of freedom needed to
describe physics in the bulk.

\emph{Optical Horizon:} Given a world-line $P(t)$ in spacetime, the
\emph{optical horizon} associated to $P(t)$ is the boundary of the
region of spacetime formed by all points $q$ connected to some point
in $P(t)$ through null geodesics. We will denote the \emph{optical
  horizon} by $\aleph$. Note that a point $q$ can be beyond the
optical horizon of $P(t)$, and yet be causally connected (by following
non-geodesic motion). This is in contrast with the \emph{domain of
  influence}~\cite{Geroch} of a point in spacetime. Also, the optical
horizon will not in general be an event horizon.

\emph{Chronologically safe region:} We call \emph{chronologically safe
  region} a region of spacetime that does not fully contain any closed
timelike curve. Notice that we don't exclude the possibility that
closed timelike curves go through a chronologically safe region, as
long as they are not fully contained in it.

Much of the present paper is concerned with the relations among these
regions for ``overrotating'' axisymmetric spacetimes. We will be
considering a comoving observer at the origin. Then the holographic
screen and the optical horizon are constructed by considering the null
geodesics that go through the world-line of this observer.
Furthermore, we will be interested in chronologically safe regions
centered around the origin and preserving the rotational symmetry in
each plane.

We can readily establish some relations between these different
surfaces.  For instance, it is obvious that the holographic screen is
inside the optical horizon. Not only that, it has to be strictly
inside: the reason is that for the metrics we consider, the expansion
of null rays starts being positive, and at the optical horizon has to
be negative. Since at the position of the holographic screen the
expansion is zero by definition, it has to be placed before reaching
the optical horizon.

Our ultimate objective is to show that for the spacetimes we study,
the induced metric on the holographic screen is free of closed
timelike curves.  To do so, we want to prove that the holographic
screens are always inside the corresponding chronologically safe
regions.

\subsection{Closed Timelike Curves}

Even though we left the direction $\tau$ noncompact, the
spacetime~\eqref{ctcmetric} has closed timelike curves.  For example,
the Killing vector $\partial_\phi$, which has periodic orbits, becomes
timelike when $\rho$ is large enough. We will now analyze the
structure of the CTCs. We will see that, as in~\cite{Boyda:2002ba},
the CTCs are ``topologically large''. This means that for a given
observer there is always a region around him that contains no CTCs.

Let us consider a comoving observer placed at $(\rho,\theta)=(0,0)$,
and moving along the $\tau$ direction.  The periodic coordinates are
given by $\phi$ and $\varphi$.  Therefore any CTC must involve these
coordinates. Moreover, in any region where the metric on the
$(\phi,\varphi)$ torus is Euclidean, there can be no CTC. So to
analyze where CTCs might occur, we have to analyze the signature of
the metric restricted to this torus.  This metric is given by
\begin{eqnarray}
  g_{(\phi,\varphi)} &=& 
    4R^2\sinh^2\frac{\rho}{2R}\biggl(1-(j^2-1)\sinh^2\frac{\rho}{2R}\biggr)
      d\phi^2 
    + 4R^2\sin^2\frac\theta2\biggl(1-j^2\sin^2\frac\theta2\biggr)
      d\varphi^2 
  \nonumber\\
    && + 8j\sqrt{j^2-1}R^2\sinh^2\frac{\rho}{2R} \sin^2\frac\theta2\;
         d\phi\, d\varphi\, . 
\end{eqnarray}
We want to know for which values of $(\rho,\theta)$ this metric is
spacelike. The signature is determined by the determinant of this
metric, as there can never be more than one timelike eigenvalue. This
determinant is easily calculated and is given by
\begin{equation}
  \det g_{(\phi,\varphi)} = 16R^4\sinh^2\frac{\rho}{2R} \sin^2\frac\theta2
    \biggl(1-(j^2-1)\sinh^2\frac{\rho}{2R}-j^2\sin^2\frac\theta2\biggr)\, . 
\end{equation}
We conclude that for 
\begin{equation}\label{saferegion}
  (j^2-1)\sinh^2\frac{\rho}{2R}+j^2\sin^2\frac\theta2 < 1\, , 
\end{equation}
the metric $g_{(\phi,\varphi)}$ is Euclidean. Hence in this region
there will not be any CTC, in other words it is a chronologically safe
region. The region is a disc in the $(\rho, \theta)$ plane centered
around the observer.

For $(\rho,\theta)$ strictly outside this chronologically safe region,
the metric $g_{(\phi,\varphi)}$ has a timelike direction. One easily
sees that there is a closed curve in the $(\phi,\varphi)$ plane, which
has the topology of a 2-torus, which is timelike everywhere. The
corresponding Killing vector will then generate a CTC\footnote{This
  can be seen as follows. The Killing vector
  $\xi_{p,q}=p\partial_\phi+q\partial_\varphi$ generates a closed
  curve for $p,q\in\Z$. Let's say that
  $\xi_{a,b}=a\partial_\phi+b\partial_\varphi$ has negative norm
  squared. Then we can approximate the slope $\frac{a}{b}$ by a
  fractional number $\frac{p}{q}$ as close as we want.  As the norm of
  $\xi_{a,b}$ has strictly negative norm squared, there must be
  $p,q\in\Z$ such that $\xi_{p,q}$ has negative norm squared.}.
Hence~\eqref{saferegion} is the largest rotationally invariant safe
region around the observer at the origin.  In the following we will
denote by $(\rho_c,\theta_c)$ the values lying on the critical surface
forming the boundary of~\eqref{saferegion}.

\subsection{Geodesics}

Our next goal will be to find the optical horizon for an observer at
the origin. To find this, we will study the geodesics of the
metric~\eqref{ctcmetric}. As we are interested in the reach of the
causal geodesics in this spacetime, we will mostly focus on the null
geodesics.

Due to the large symmetry of the spacetime, the geodesic equations are
easily solved taking advantage of the corresponding conserved charges.
We will mainly make use of the conserved charges corresponding to the
three Killing vectors $\xi_0$, $\xi_3$ and $\xi_6$.  These are given
by the energy and two angular momenta:
\begin{eqnarray}
\label{pitau}
  E &=& \dot\tau + 2jR\sinh^2\frac{\rho}{2R}\;\dot\phi 
        - 2\sqrt{j^2-1}R\sin^2\frac\theta2\;\dot\varphi\, , \\
\label{piphi}
  J_\phi &=& R^2\sinh^2\frac{\rho}{R}\;\dot\phi 
             - jRE\cosh\frac{\rho}{R}\, , \\
\label{pivarphi}
  J_\varphi &=& R^2\sin^2\theta\;\dot\varphi 
                - \sqrt{j^2-1}RE\cos\theta\, .
\end{eqnarray}
These conserved momenta can be used to find the solutions to the
motion of the angles $\phi$ and $\varphi$ and for $\tau$.  Apart from
the conserved quantities associated to the Killing vectors, we have
the conserved first integral given by
\begin{equation}\label{integral}
  -m^2  = \dot\rho^2 + R^2\sinh^2\frac{\rho}{R}\;\dot\phi^2 
          + R^2\dot\theta^2  + R^2\sin^2\theta\;\dot\varphi^2 - E^2\, .
\end{equation}
Here $m^2$ is positive, negative, or zero for timelike, spacelike or
null geodesics, respectively.

Because of homogeneity and rotational symmetry in the $\phi$ and
$\varphi$ angles, we can, without loss of generality, restrict to the
analysis of the geodesics at constant $\rho$ and $\theta$.  Note that
for such geodesics $\dot\phi$, $\dot\varphi$ and $\dot\tau$ are
constant, so their motion is linear in the affine parameter. All other
geodesics can be found by acting on these by the isometries.  Note
that we have precisely enough parameters in this class of geodesics
to, together with the isometries, generate the full set of
them\footnote{There are three parameters $E$, $J_\phi$ and
  $J_\varphi$, that determine the geodesics centered around the
  origin.  Shifting them to go through the origin will introduce two
  angular parameters, describing the orientation in the two
  independent planes. According to the equations of motion, the
  geodesics through the origin should form a five parameter family of
  solutions, labeled by five initial velocities. }.

To find the position in the $(\rho,\theta)$ plane of such geodesics we
will have to consider the equations of motion for these coordinates.
For $\rho$ the equation can be written as
\begin{equation}\label{rhoeom}
  \ddot\rho = \frac{(J_\phi\cosh\frac{\rho}{R}+jRE)
                    (J_\phi+jRE\cosh\frac{\rho}{R})}%
                   {R^3\sinh^3\frac{\rho}{R}}\, ,
\end{equation}
where we used~\eqref{piphi}.
Similarly the equation of motion for $\theta$ can be written 
\begin{equation}\label{thetaeom}
  \ddot\theta = \frac{(J_\varphi\cos\theta+\sqrt{j^2-1}RE)
                      (J_\varphi+\sqrt{j^2-1}RE\cos\theta)}%
                     {R^4\sin^3\theta}\, .
\end{equation}

We want to solve the geodesics for fixed $\rho$ and $\theta$.
Therefore we have to set the left-hand side of~\eqref{rhoeom}
and~\eqref{thetaeom} to zero.  This allows us to solve for $\rho$ and
$\theta$ in terms of the conserved charges.  From the equations of
motion for $\rho$~\eqref{rhoeom} we find two solutions for $\rho$,
satisfying
\begin{equation}\label{rhosol}
  (i)\quad \cosh\frac{\rho}{R} = -\frac{jRE}{J_\phi}\, ,
  \quad\mbox{or}\quad
  (ii)\quad \cosh\frac{\rho}{R} = -\frac{J_\phi}{jRE}\, . 
\end{equation}
Similarly~\eqref{thetaeom} gives two solutions for $\theta$ 
\begin{equation}\label{thetasol}
  (i)\quad \cos\theta = -\frac{\sqrt{j^2-1}RE}{J_\varphi}\, ,
  \quad\mbox{or}\quad 
  (ii)\quad \cos\theta = -\frac{J_\varphi}{\sqrt{j^2-1}RE}\, .
\end{equation}

Taking the second solutions both in~\eqref{rhosol} and
in~\eqref{thetasol} will lead to the trivial timelike geodesic at
$\rho=\theta=0$, followed by the observer in the origin. The other
three combinations may lead to null geodesics, and will now be
discussed in turn.

\paragraph{Null Geodesics in the Hyperbolic Plane}

We first discuss the combination of solution~($i$) in~\eqref{rhosol}
and~($ii$) in~\eqref{thetasol}. From~\eqref{pivarphi} we find that
$\dot\varphi=0$, so the geodesic will stay at a fixed point on the
sphere and hence moves only in the hyperbolic plane and the $\tau$
direction.  Because of this, the geodesics will be exactly those
of~\cite{Drukker:2003mg} for the case of the hyperbolic plane.  We may
just as well assume that $\theta=0$.  The first
integral~\eqref{integral} for null geodesics ($m=0$) gives
\begin{equation}
  J_\phi^2=(j^2-1)R^2E^2\, . 
\end{equation}
Therefore the geodesic is located at $\rho_g$ determined by 
\begin{equation}
  \cosh\frac{\rho_g}{R} = \frac{j}{\sqrt{j^2-1}}\, . 
\end{equation}
Comparing to~\eqref{saferegion} we see that the projection of the
geodesic to the $(\rho,\phi)$ plane lies completely inside the
chronologically safe region. Even better, the homogeneity of the
solution implies a relation between the radius of the null geodesics
and the radius of the optical horizon: the projection of the geodesic
on the hyperbolic plane describes a circle centered around the origin
$\rho=0$.  Using the $\SL(2)$ symmetry we can move the geodesic to
find other geodesics whose projection moves only in the hyperbolic
plane. Of particular importance to us are the geodesics that pass
through the origin $\rho=0$. The projection of this geodesic to the
$(\rho,\phi)$ plane is a (deformed) circle. The maximal value for
$\rho$ is $\rho_c = 2\rho_g$. This relation between the radius of the
projected null geodesics and the critical radius was found
in~\cite{Drukker:2003mg} for the G\"odel type solutions with a single
hyperbolic or flat plane, and the origin is the same: the spacetime
homogeneity of the solution.  Note also that this is precisely the
radius of the chronologically safe region in the hyperbolic plane.

\paragraph{Null Geodesics in the Sphere}

Next we discuss the combination of solution~($ii$) in~\eqref{rhosol}
and~($i$) in~\eqref{thetasol}. From~\eqref{piphi} we find that
$\dot\phi=0$, so the geodesic will stay at a fixed point on the
hyperbolic plane and hence moves only in the sphere and the $\tau$
direction. We may just as well assume that $\rho=0$.  The first
integral~\eqref{integral} for null geodesics $m=0$ gives
\begin{equation}
  J_\varphi^2=j^2R^2E^2\, . 
\end{equation}
Hence the geodesic is located at $\theta_g$ determined by 
\begin{equation}
  \cos\theta_g = \frac{\sqrt{j^2-1}}{j}\, . 
\end{equation}
Again, we see that the projection of the geodesic to the
$(\theta,\varphi)$ plane lies completely inside the chronologically
safe region, and that the optical horizon is at $\theta_c=2\theta _g$.

\paragraph{Generic Null Geodesics}

We now discuss the combination of solutions~($i$) for
both~\eqref{rhosol} and~\eqref{thetasol}.  From the equations above we
find that the first integral~\eqref{integral} for null geodesics gives
\begin{equation}
  J_\phi^2 = J_\varphi^2\, . 
\end{equation}
We therefore find that these geodesics move both in the hyperbolic
plane and the sphere.  These null geodesics will be located at
$(\rho_g,\theta_g)$ which are determined by the first solutions
in~\eqref{rhosol} and~\eqref{thetasol}.  We derive from this and the
relation between $J_\phi$ and $J_\varphi$ that these points satisfy
\begin{equation}\label{geodpos}
  (j^2-1)\sinh^2\frac{\rho_g}{R} + j^2\sin^2\theta_g 
  = (j^2-1)\Bigl(\cosh^2\frac{\rho_g}{R}-1\Bigr)+j^2(1-\cos^2\theta_g)
  = 1\, . 
\end{equation}

Comparing this to~\eqref{saferegion}, we see that these geodesics lie
completely inside the safe region for the observer at the origin.
However we can say more, making use of the homogeneity of the
spacetime.  We have analyzed the geodesics which are centered around
$(\rho,\theta)=(0,0)$.  From this we can find all other solutions by
acting with the $\SU(2)\times\SL(2)$ isometries.  Note that the
geodesics project to circles in the $(\rho, \phi)$ and the
$(\theta,\varphi)$ planes. Acting with the isometries we can shift the
null geodesics such that they go through the origin at
$(\rho,\theta)=(0,0)$.  If we project this geodesic to the
$(\rho,\theta,\phi,\varphi)$ space we can easily see that the point
furthest away from the origin will be at
$(\rho,\theta)=(2\rho_g,2\theta_g)$. Comparing~\eqref{geodpos}
and~\eqref{saferegion} we see that this point lies exactly on the
boundary of the chronologically safe region.

\begin{figure}[ht]
  \begin{center}
    \includegraphics{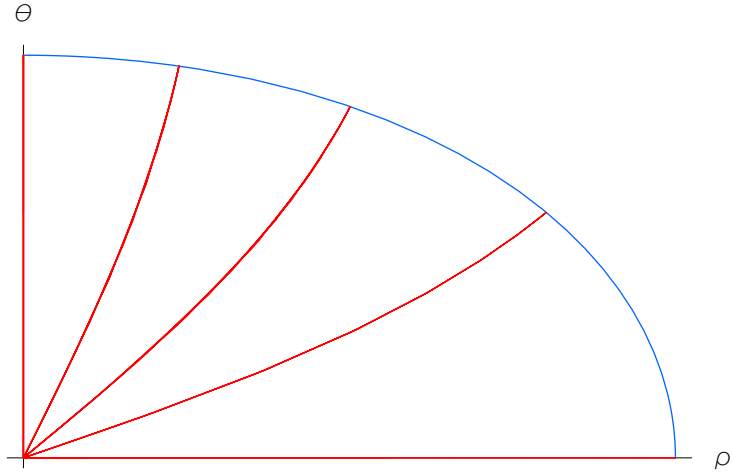}
    \caption{Projection of some null geodesics to the $(\rho,\theta)$ 
      plane for the near horizon BMPV solution at $j=1.1$. The quarter
      ellipse is the boundary of the chronologically safe region,
      which coincides with the optical horizon $\aleph$.}
    \label{fig:nullgeod}
  \end{center}
\end{figure}

In \figref{fig:nullgeod} we sketched the projection to the
$(\rho,\theta)$ plane of some null geodesics going through the origin.
The analysis of the geodesics above can be summarized by saying that
the optical horizon $\aleph$ is exactly the same as the boundary of the
chronologically safe region.

\subsection{Global Issues and NUT Singularity}

The metric~\eqref{ctcmetric} interpreted as in this section has a
problem when one considers it globally. As the coordinate $\varphi$ is
periodic with period $2\pi$, and the length of the coordinate vector
$\partial_\varphi$ vanishes at $\theta=0,\pi$, there are potential
singularities at these loci. With the metric as written
in~\eqref{ctcmetric} there is no singularity at $\theta=0$, because
$\|\partial_\varphi\|\sim\theta R$. However at $\theta=\pi$ the norm
of $\partial_\varphi$ does not vanish, and therefore there will be a
singularity. This singularity is of Taub-NUT type. Indeed, forgetting
about the hyperbolic plane, we have exactly the angular part of the
Lorentzian Taub-NUT spacetime. It is well known~\cite{Misner} that the
singularity at $\theta=\pi$ can be removed if one takes the coordinate
$\tau$ periodic with period given by $4\pi\sqrt{j^2-1}R$ (the NUT
charge). Indeed, replacing $\tau$ by
$\tau'=\tau-2\sqrt{j^2-1}R\varphi$ leads to good local coordinates
near $\theta=\pi$. Note that this coordinate transformation is only
allowed if $\tau$ has the correct periodicity.

This traditional point of view leads to an immediate problem with the
proposal of holographic chronology protection of~\cite{Boyda:2002ba}.
Since $\partial_\tau$ generates a timelike geodesic, if we make $\tau$
periodic, the induced metric on the holographic screen will have CTCs.

However, one can take a different point of view, which also interplays
nicely with the holographic chronology protection.  We notice that if
we keep the NUT singularity at $\theta=\pi$, the holographic screen
enclosing the observer at the origin is still causal and nonsingular.
Indeed, the point $\theta=\pi$ lies strictly outside the
chronologically safe region.  Taking the point of view
of~\cite{Boyda:2002ba}, the CTCs as well as the singularities should
not affect the holographic description of the physics accessible to
our observer. We stress that this is quite a strong statement, since
there is no event horizon surrounding the singularity. Hence,
classically there are still causal paths connecting the observer to
the singularity. A more palatable option is that these metrics only
show up in a finite region of spacetime, and are patched to exterior
metrics without CTCs, as in~\cite{Drukker:2003sc}.

\section{CTCs and Optical Horizons in Axisymmetric Spaces}
\label{sec:ctcgeneral}

In the last section we saw that for the overrotating near horizon
BMPV metric, the optical horizon is coincident with the boundary of
the chronologically safe region.  In other words, the observer has no
access through geodesic motion to the region where CTCs centered
around him appear.

In this section we will discuss to which extent this feature is
present in more general stationary axisymmetric space-times, in an
arbitrary number of planes.

\subsection{Axisymmetric Metrics}

We consider a metric which is stationary and in addition is
rotationally symmetric. More concretely, we will have a time
coordinate $t$ and angular coordinates $\phi^i$, such that the vectors
$\partial_t$ and $\partial_{\phi^i}$ are Killing. Furthermore we have
several ``radial'' coordinates $\rho^a$. Our ansatz for the metric
will be
\begin{equation}\label{cylmetric}
  ds^2 = -k^2\Bigl(dt + A_i(d\phi^i+C^i_a\,d\rho^a) + B_a\,d\rho^a\Bigr)^2 
         + g_{ij}(d\phi^i+C^i_a\,d\rho^a)(d\phi^j+C^j_b\,d\rho^b) 
         + h_{ab}\,d\rho^a\,d\rho^b\, ,
\end{equation}
where all functions may depend on the coordinates $\rho^a$.  $k$,
$g_{ij}$ and $h_{ab}$ are positive definite. Furthermore we assume the
metric to be nonsingular. More precisely, we will assume that, after
an appropriate coordinate transformation for $\rho$, near the origin
$\rho=0$ $k$ is of order one, $g_{ij}$ is of order $\rho^2$ and $A_i$
is of order smaller than $\rho$.  Note that not all $\rho^a$ have to
be radial coordinates.  However we will assume that the metric
restricted to the $\rho$ plane is always Euclidean. This means that
the $B_a$ should be bounded with respect to the metric $h_{ab}$.

Metrics of this form include the Van Stockum solution~\cite{Stockum},
the overrotating supertube solution~\cite{Emparan:2001ux}, and many
$(2n+1)$-dimensional G\"odel universes formed by ``rotation'' in
products of hyperbolic, spherical, or flat planes~\cite{Reboucas:hn,
  Harmark:2003ud}. The different values for the metric $g_{ij}$ and
the connection $A_i$ for these situations are given schematically in
\tbref{tab:godelplane}.

\begin{table}[htbp]
  \begin{center}
    \begin{tabular}{llll}
      plane      & $g$           & $A$ \\\hline
      flat       & $\rho^2$      & $\frac12a\rho^2$ \\
      hyperbolic & $\sinh^2\rho$ & $2a\sinh^2\frac\rho2$ \\
      sphere     & $\sin^2\rho$  & $2a\sin^2\frac\rho2$ 
    \end{tabular}
    \caption{Metrics and connections on G\"odel planes.}
    \label{tab:godelplane}
  \end{center}
\end{table}

\subsection{Closed Timelike Curves}

Let us first identify the closed timelike curves. We will be
interested in an observer located at the origin $\rho=0$. By our
assumption closed timelike curves necessarily involve nontrivial
motion in the $\phi$ plane. Therefore to find the chronologically safe
region we should consider the metric on the $\phi$ plane. As argued in
the last section, in any region in the $\rho$ plane where this metric
is positive definite there can not be a CTC.

The metric on the $\phi$ plane is given by 
\begin{equation}
   g_{(\phi)} = (g_{ij}-k^2A_iA_j)\,d\phi^i\,d\phi^j\, . 
\end{equation}
Because there is at most one time direction, the signature of this
metric is completely determined by the sign of its determinant. This
determinant is easily calculated,
\begin{equation}
  \det g_{(\phi)} = (1-k^2g^{ij}A_iA_j)\det g_{ij}\, . 
\end{equation}
We conclude that in the region determined by 
\begin{equation}\label{saferegion2}
  k^2g^{ij}A_iA_j < 1\, ,
\end{equation}
the metric on the $\phi$ plane is Euclidean and therefore does not
contain any closed timelike curve. Notice that this region encloses
the observer at the origin.  Moreover, strictly outside this region,
where the determinant is negative, one finds closed curves moving only
in the $\phi$ plane which are everywhere timelike.

\subsection{Geodesics}

We will now analyze the geodesics for a metric of the form
\eqref{cylmetric}.  The symmetries generated by the Killing vectors
$\partial_t$ and $\partial_{\phi^i}$ give rise to conserved charges
\begin{eqnarray}
  E   &=& k^2(\dot t + A_i(\dot\phi^i+C^i_a\dot\rho^a) + B_a\dot\rho^a)\, , \\
  J_i &=& g_{ij}(\dot\phi^j+C^j_a\dot\rho^a)-EA_i.
\end{eqnarray}
The momenta conjugate to the coordinates $\rho^a$ are given by 
\begin{equation}
  \pi_a = h_{ab}\dot\rho^b + C^i_aJ_i - EB_a\, .
\end{equation}
The Hamiltonian for the geodesic flow, $H=g^{\mu \nu}\pi_\mu \pi_\nu$, 
is in this case 
\begin{eqnarray}
  H &=& -\frac{E^2}{2k^2} + \frac12 g^{ij}(J_i+EA_i)(J_j+EA_j) 
  \nonumber\\
  && + \frac12 h^{ab}(\pi_a+EB_a-J_iC^i_a)(\pi_b+EB_b-J_jC^j_b)\, . 
\end{eqnarray}

We are interested in the null geodesics that pass through the
worldline of the observer at the origin. We find from the form of the
momenta $J_i$ and our assumptions on the metric components near
$\rho=0$ that such geodesics must necessarily have $J_i=0$.  This is
natural, as they can not have angular momenta when they pass the
origin. For these geodesics the form of the Hamiltonian $H$,
simplifies considerably. Furthermore, for null geodesics the
Hamiltonian constraint reads $H=0$, so all in all,
\begin{equation}\label{hamilcon}
  H = -\frac{E^2}{2k^2}(1-k^2g^{ij}A_iA_j) 
      + \frac12h^{ab}(\pi_a+EB_a)(\pi_b+EB_b) 
    = 0\, . 
\end{equation}
We recognize in the first term, the effective potential, the same
factor we saw in the analysis of the closed timelike curves.  Because
the last term is non-negative, we conclude that any geodesic passing
through $\rho=0$ will remain in the chronologically safe region
\eqref{saferegion2}.

\subsection{Generic Location of the Optical Horizon}

We now want to compare the location of the optical horizon with the
boundary of the chronologically safe region. We just showed that for
these spacetimes, the optical horizon never reaches beyond the
chronologically safe region. In what follows we will argue that
typically the optical horizon coincides with the boundary of the
chronologically safe region (in a sense to be clarified below),
although we will also show that in some cases the optical horizon is
strictly inside such region.

Let us, as a simple example, consider a $(2n+1)$-dimensional spacetime
which is a set of $n$ flat G\"odel planes, i.e., the metric is given
by
\begin{equation}
  ds^2 = -\Bigl(dt+\sum_{i=1}^n\omega_ir_i^2d\phi_i\Bigr)^2
         + \sum_{i=1}^n(r_i^2d\phi_i^2+dr_i^2)\, ,
\end{equation}
with $\omega_i$ some constants. The chronologically safe region 
is determined by the ball 
\begin{equation}\label{csrgodel}
  \sum_{i=1}^n\omega_i^2r_i^2< 1\, . 
\end{equation}

The null geodesics passing through $r=0$ for this spacetime are easily
found using the discussion above.  Setting $E=1$, they are given by
\begin{equation}
  r_i(\lambda) = R_i\sin\omega_i\lambda\, ,
\end{equation}
with 
\begin{equation}
  \sum_{i=1}^n\omega_i^2R_i^2=1\, . 
\end{equation}

The angles $\phi^i$ will depend linearly on $\lambda$. It follows from
the previous equations that the null geodesics never escape the
chronologically safe region. Furthermore, we recognize the projection
of the null geodesics as a Lissajous figure, which will be a closed
curve iff all the ratios $w_i/w_j$ are rational.

We now discuss in turn the possible cases, depending on the values of
$w_i/w_j$. Assume first that all these ratios are rational. Then, we
can still distinguish two possibilities: either all the ratios are
given by fractions of odd integers, or they are not. In the first
case,

\begin{equation}
  \frac{w_i}{w_j} = \frac{2n_i+1}{2n_j+1}\, ,\qquad 
  \forall \; i,j\, ,\qquad n_i \in \Z\, ,
\end{equation}
it follows that there is a value of $\lambda $ for which all
$r_i(\lambda)= R_i$, (or equivalently, all $\dot r_i =0$
simultaneously), and the projection of the null geodesic manages to
touch the boundary of the chronologically safe region, before focusing
back towards the origin.  We conclude that in this case the optical
horizon coincides with the boundary of the chronologically safe
region.

Although we carried out this analysis for metrics with flat G\"odel
planes, it is true more generally. For instance, for the near horizon
BMPV metric discussed in the last section, we already showed that the
frequencies of rotation in the hyperbolic plane and the sphere are
equal, for the generic null geodesic.  This is reflected in
\figref{fig:nullgeod}, where we see that for the near horizon BMPV
metric, the projection of the geodesics to the $\rho$ plane collapses,
that is it follows exactly the same path going away from the origin as
when it comes back. This finetuning of frequencies is actually common
among the supersymmetric solutions discussed in the literature
\cite{Harmark:2003ud}.

Next, we consider the case when the ratios $w_i/w_j$ are still all
rational, but now some are given by a fraction involving an even and
an odd integer

\begin{equation}
 \frac{w_i}{w_j} = \frac{2n_i}{2n_j+1}\, ,\qquad
  \mbox{for some } i,j\, ,\qquad n_i \in \Z\, .
\end{equation}

In this situation, the null geodesics can reach the boundary of the
chronologically safe region only if they don't have momentum in the
$r_i$ directions associated to an even integer, so in general the
optical horizon touches the boundary chronologically safe region, but
does not coincide with it.

\begin{figure}[htbp]
  \begin{center}
    \includegraphics{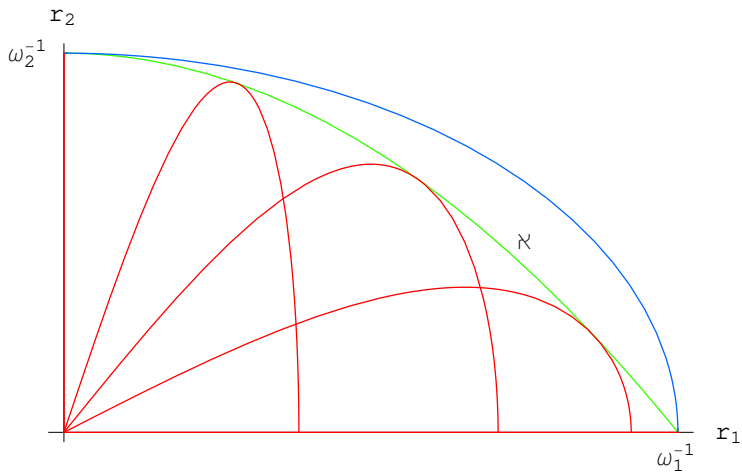}
    \caption{Projection of some null geodesics to the $r_1,r_2$-plane 
      for the flat G\"odel solution in 5 dimensions for
      $\omega_2=2\omega_1$.  The outer curve is the boundary of the
      chronologically safe region, the inner curve $\aleph$ is the
      optical horizon, which is the hull of the projected null
      geodesics through the origin. }
    \label{fig:nullgeod3}
  \end{center}
\end{figure}

Perhaps an example will clarify this. Consider the metric for two flat
G\"odel planes with $\omega_2/\omega_1=2$. The projections of some
null geodesics to the $(r_1,r_2)$ plane passing through the origin are
drawn in \figref{fig:nullgeod3}. The figure already suggests that, as
we just discussed, these null geodesics will not all come close to the
boundary of the safe region, indicated in the figure by the enclosing
quarter ellipse.  In this simple example we can do even better, and
find explicitly the optical horizon. To do so, notice that
\begin{eqnarray}
  1-\omega _1^2r_1^2(\lambda)-2\omega_1 r_2(\lambda) 
  &=& 1 - \omega_1^2 R_1^2\sin^2\omega_1\lambda 
      - 2\omega_1 R_2\sin2\omega_1\lambda 
  \nonumber\\
  &=& (2\omega_1R_2\sin\omega_1\lambda-\cos\omega_1\lambda)^2 
    \geq 0\, , 
\nonumber
\end{eqnarray}
and moreover the bound is saturated for $\cot\lambda=2\omega_1R_2$.
This implies that the optical horizon $\aleph$ is given by the surface
at
\begin{equation}
  \omega_1^2r_1^2+2\omega_1 r_2 = 1\, ,
\end{equation}
which lies strictly inside the chronologically safe region, given by
\eqref{csrgodel}.

\begin{figure}[htbp]
  \begin{center}
    \includegraphics{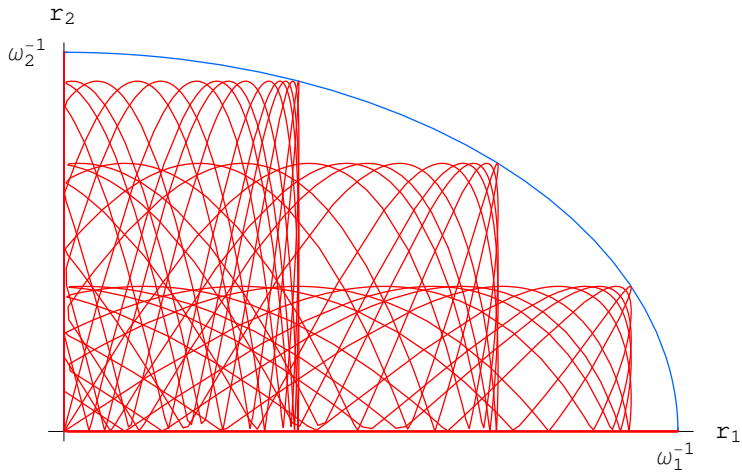}
    \caption{Projection of some null geodesics to the $r_1,r_2$-plane 
      for the flat G\"odel solution in 5 dimensions for
      $\omega_2/\omega_1=2.61799\ldots$.  The outer curve is the
      boundary of the chronologically safe region, as well as the
      optical horizon $\aleph$.  The geodesics are followed only for a
      finite time; a full null geodesic would densely fill a
      rectangular region.}
    \label{fig:nullgeod2}
  \end{center}
\end{figure}

Finally, let's consider the generic case, when some or all of the
ratios $w_i/w_j$ are not rational. Now the projections of the
geodesics are no longer closed curves. Such a situation is sketched in
\figref{fig:nullgeod2}.  As suggested by the figure, the geodesics
will densely fill a rectangular region bounded by $r_i=R_i$. A corner
of this rectangle will exactly lie on the boundary of the
chronologically safe region. So even though the geodesics will never
reach this boundary, they will come arbitrarily close to it. This can
be argued more precisely from the form of the solutions of the
geodesics in the $r$ plane. Indeed, when the $\omega_i$ have
non-rational quotients, we can make the $\dot r_i(\lambda)$
simultaneously arbitrarily small, which is equivalent, according
to~\eqref{hamilcon}, to moving arbitrarily close to the boundary of
the safe region.

For more general metrics of the form~\eqref{cylmetric}, we can argue
that generically the null geodesics will come arbitrarily close to the
boundary of the safe region. We saw that the motion of the geodesics
in the $\rho$ plane was governed by a Hamiltonian flow with total zero
energy.  Furthermore, the chronologically safe region is a cavity
around $\rho=0$ bounded by the hypersurface of zero effective
potential.  We can now invoke the Poincar\'e recurrence theorem. It
states in particular that for generic parameters in the potential, any
integral curve of this motion will densely fill the phase space region
of constant energy. Here we assumed that the cavity is in fact
compact.  This part of the phase space will project to the whole
cavity inside zero effective potential. This implies that any null
geodesic will come arbitrarily close to the boundary of the safe
region, and therefore the optical horizon will coincide with this
boundary.

To summarize, for metrics with flat G\"odel planes, the optical
horizon coincides with the boundary of the chronologically safe
region, unless some frequencies satisfy $w_i/w_j=2n_i/(2n_j+1)$.

Note that in the example of flat G\"odel planes in this section the
individual geodesics did not fill the whole cavity, but rather a
rectangular region fitting just inside it. Of course the form of the
metric was certainly not generic, as the radial dependence of the
various functions was taken to be particularly simple.

\section{Discussion and Conclusions}
\label{sec:concl}

We have studied in detail an example of a spacetime homogeneous 5d
supergravity solution with closed timelike curves, the near horizon of
an overrotating BMPV black hole. As discussed in previous sections,
this solution can be thought of as two G\"odel type planes (one
hyperbolic, one spherical) related to the 4d G\"odel type solutions
presented in~\cite{Reboucas:hn} and studied in~\cite{Drukker:2003mg,
  Israel:2003cx}. As a result, the qualitative pattern of the regions
with CTCs and optical horizon is quite similar. Some of our results,
like the relation between null geodesics and optical horizon, are a
direct consequence of the spacetime homogeneity of the solution, so we
expect them to be true in arbitrary spacetime homogeneous solutions.

As a welcome spin-off of our study, we have brought to a full
conclusion the classification of maximally supersymmetric solutions in
five dimensional minimal supergravity. Namely, we found that the three
unidentified solutions of~\cite{Gauntlett:2002nw} belong to the
NH-BMPV family, so the five solutions listed in the introduction are
the full set of maximally supersymmetric backgrounds of this theory.
Furthermore we showed explicitly how they arise from reductions of
$AdS_3\times S^3$, and how the different reductions determine their
causal structures. Finally, we also elucidated the relation between
$AdS_3\times S^2$ and the NH-BMPV family, arguing than the former does
not belong to the NH-BMPV family.

We also considered more general axisymmetric metrics with closed
timelike curves, not necessarily homogeneous. Within the class of
metrics discussed, we showed that the null geodesics passing through
the origin never escape the chronologically safe region. We expect
this to be true in much more generality, and it would be interesting
to generalize our proof. In particular, our arguments show that the
observation of~\cite{Boyda:2002ba} that holographic screens are inside
the chronologically safe region for G\"odel type solutions applies to
a much wider class of metrics.

The physical relevance of such observation depends of course on the
possibility of realizing such metrics. Within the context of string
theory, probe computations~\cite{Emparan:2001ux, Dyson:2003zn,
  Drukker:2003sc} suggest in particular cases that ten dimensional
solutions with closed timelike curves can't be built starting from
flat space. It would be important to understand the generality of this
assertion.

\section*{Acknowlegements}

We would like to thank Jos\'e Figueroa-O'Farrill, Veronika Hubeny,
Patrick Meessen and Tom\'as Ort\'{\i}n, for useful conversations. BF
would also like to thank the ITP at Stanford University and the KITP
at UCSB for hospitality.  The work of BF was supported by a Marie
Curie Fellowship. The work of E.L.-T.  was supported in part by by the
Spanish grant BFM2003-01090. The work of C.H. was partly supported by
a Koshland Scholarship.  In addition, this work was supported by the
Israel-U.S. Binational Science Foundation, by the ISF Centers of
Excellence program, by the European network HPRN-CT-2000-00122 and by
Minerva.

\appendix

\section{Hopf Fibrations of $S^3$ and $AdS_3$}
\label{app:hopf}

In this appendix we will write down the explicit Hopf fibrations of 
$S^3$ and especially $AdS_3$.

\subsection{Hopf Fibration of $S^3$}

We start with the well known Hopf fibration of $S^3$, identified with
the group manifold $\SU(2)$. We use the anti-Hermitian basis for
$\su(2)$ given by $\hat\tau^a=\frac{i}{2}\sigma^a$, where $\sigma^a$
are the standard Pauli spin matrices. We then parametrize the group
elements $g\in\SU(2)$ as
\begin{equation}
  g = \e^{\phi\hat\tau^3}\e^{\theta\hat\tau^2}\e^{\psi\hat\tau^3}\, .
\end{equation}
The left-invariant one-forms $\omega^a$ can be found from 
\begin{equation}
  g\inv dg = \omega^a\hat\tau^a\, . 
\end{equation}
We find 
\begin{equation}
  \left\{
  \begin{array}{r@{\;}c@{\;}l}
    \omega^1 &=& -\sin\psi\, d\theta + \cos\psi\sin\theta\, d\phi\, , \\
    \omega^2 &=& \cos\psi\, d\theta + \sin\psi\sin\theta\, d\phi\, , \\
    \omega^3 &=& d\psi + \cos\theta\, d\phi\, .
  \end{array}
  \right.
\end{equation}

The metric is given by $-\tr(g\inv dg)^2=\frac14\sum_a(\omega^a)^2$,
where $\tr$ is a trace normalized such that $\tr\one=1$.  This leads
to the metric
\begin{equation}
  d\Omega_{(3)}^2 = \frac14 d\theta^2 + \frac14 \sin^2\theta\;d\phi^2
                    + \frac14 (d\psi+\cos\theta\;d\phi)^2\, . 
\end{equation}

The Killing vectors for the left action of $\SU(2)$ are given by
\begin{equation}
  \left\{
  \begin{array}{r@{\;}c@{\;}l}
    \xi^L_1 &=& \displaystyle 
                \sin\phi\;\partial_\theta + \cos\phi\cot\theta\;\partial_\phi 
                - \cos\phi\cosec\theta\;\partial_\psi\, , \\
    \xi^L_2 &=& \displaystyle 
                \cos\phi\;\partial_\theta - \sin\phi\cot\theta\;\partial_\phi 
                + \sin\phi\cosec\theta\;\partial_\psi\, , \\
    \xi^L_3 &=& \partial_\phi\, ,
  \end{array}
  \right.
\end{equation}
while the right action of $\SU(2)$ is generated by the Killing vectors  
\begin{equation}
  \left\{
  \begin{array}{r@{\;}c@{\;}l}
    \xi^R_1 &=& \displaystyle 
                -\sin\psi\;\partial_\theta - \cos\psi\cot\theta\;\partial_\psi 
                + \cos\psi\cosec\theta\;\partial_\phi\, , \\
    \xi^R_2 &=& \displaystyle 
                \cos\psi\;\partial_\theta - \sin\psi\cot\theta\;\partial_\psi 
                + \sin\phi\cosec\theta\;\partial_\phi\, , \\
    \xi^R_3 &=& \partial_\psi\, . 
  \end{array}
  \right.
\end{equation}
These Killing vectors are normalized such that
$\xi^R_i\cdot\omega_L^j=\xi^L_i\cdot\omega_R^j=\delta_i^j$.

\subsection{Spacelike Hopf Fibration of $AdS_3$}

For $AdS_3$ we can write down Hopf fibrations very similar as for the
one over $S^3$. For this we identify $AdS_3$ with
$\widetilde{\SL}(2)$.  We will use the basis of $\lsl(2)$ given by the
real matrices $\hat\tau^1=\frac{1}{2}\sigma^1$,
$\hat\tau^2=\frac{i}{2}\sigma^2$, and
$\hat\tau^3=\frac{1}{2}\sigma^3$.

In the $AdS_3$ case there are essentially three different Hopf
fibrations, depending on whether the Hopf fiber is in a spacelike,
timelike, or lightlike direction. We will first study the spacelike
case.

For the spacelike Hopf fibration we take the following parametrization
of $g\in\SL(2)$,
\begin{equation}
  g = \e^{\tau\hat\tau^2}\e^{\rho\hat\tau^1}\e^{\psi\hat\tau^3}. 
\end{equation}
The Hopf fibration is given by the right action of the hyperbolic
one-parameter subgroup generated by $\hat\tau^3$.\footnote{Here and
  below we will use $\psi$ for the coordinate along the Hopf fiber.}
The left-invariant one-forms $\omega^a$ are determined in the same way
as for $S^3$, and in this case are given by
\begin{equation}
  \left\{
  \begin{array}{r@{\;}c@{\;}l}
    \omega^1 &=& \cosh\psi\, d\rho - \sinh\psi\cosh\rho\, d\tau\, , \\
    \omega^2 &=& -\sinh\psi\, d\rho + \cosh\psi\cosh\rho\, d\tau\, , \\
    \omega^3 &=& d\psi + \sinh\rho\, d\tau\, .
  \end{array}
  \right.
\end{equation}
The metric will now be given by $4\tr(g\inv dg)^2=\sum_a(\omega^a)^2$.
It can then be written
\begin{equation}\label{hopfspace}
   ds^2 = -\cosh^2\rho\, d\tau^2+d\rho^2+(d\psi+\sinh\rho\, d\tau)^2\, . 
\end{equation}

\subsection{Timelike Hopf Fibration of $AdS_3$}

For the timelike Hopf fibration we take the following parametrization
of $\SL(2)$,
\begin{equation}
g = \e^{\tau\hat\tau^3}\e^{\rho\hat\tau^1}\e^{\psi\hat\tau^2}.
\end{equation}
The Hopf fibration is given by the right action of the hyperbolic
one-parameter subgroup generated by $\hat\tau^2$.  The left-invariant
one-forms $\omega^a$ are given by
\begin{equation}
  \left\{
  \begin{array}{r@{\;}c@{\;}l}
    \omega^1 &=& \cos\psi\, d\rho + \sin\psi\cosh\rho\, d\tau\, , \\
    \omega^2 &=& d\psi + \sinh\rho\, d\tau\, , \\
    \omega^3 &=& -\sin\psi\, d\rho + \cos\psi\cosh\rho\, d\tau\, .
  \end{array}
  \right.
\end{equation}
The metric can then be written
\begin{equation}\label{hopftime}
  ds^2 = \cosh^2\rho\, d\tau^2+d\rho^2-(d\psi+\sinh\rho\, d\tau)^2.
\end{equation}

\subsection{Lightlike Hopf Fibration of $AdS_3$}

For the lightlike Hopf fibration we need yet another parametrization
of $\SL(2)$. Let us introduce parabolic generators $\hat\tau^\pm =
\hat\tau^1\pm\hat\tau^2$ of $\lsl(2)$.  We parametrize $\SL(2)$ by
\begin{equation}
  g = \e^{\tau\hat\tau^-}\e^{\rho\hat\tau^3}\e^{\psi\hat\tau^+}.
\end{equation}
The Hopf fibration is given by the right action of the parabolic
one-parameter subgroup generated by $\hat\tau^+$.  The left-invariant
one-forms $\omega^a$ are given by
\begin{equation}
  \left\{
  \begin{array}{r@{\;}c@{\;}l}
    \omega^+ &=& d\psi + \psi\, d\rho - \psi^2\e^{\rho} d\tau\, , \\
    \omega^- &=& \e^{\rho} d\tau\, , \\
    \omega^3 &=& d\rho - 2\psi\e^{\rho} d\tau\, .
  \end{array}
  \right.
\end{equation}
Therefore the metric can be written 
\begin{equation}\label{hopfnull}
  ds^2 = (\omega^3)^2 +4\omega^+\omega^- = d\rho^2+4\e^{\rho}d\tau\, d\psi\, .
\end{equation}

\subsection{Alternative Hopf Fibrations of $AdS_3$}

there are alternative parametrizations of the group $\SL(2)$, which
are actually somewhat closer to the Hopf fibration of the sphere.  For
the spacelike Hopf fibration we write
\begin{equation}
  g = \e^{\tau\hat\tau^3}\e^{\rho\hat\tau^1}\e^{\psi\hat\tau^3}\, . 
\end{equation}
The Hopf fibration can be seen as the right action by the elliptic
1-parameter subgroup of $\widetilde{\SL}(2)$ generated by
$\hat\tau^3$.  The left-invariant one-forms $\omega^a$ are given by
\begin{equation}
  \left\{
  \begin{array}{r@{\;}c@{\;}l}
    \omega^1 &=& \cosh\psi\, d\rho - \sinh\psi\sinh\rho\, d\tau\, , \\
    \omega^2 &=& -\sinh\psi\, d\rho + \cosh\psi\sinh\rho\, d\tau\, , \\
    \omega^3 &=& d\psi + \cosh\rho\, d\tau\, .
  \end{array}
  \right.
\end{equation}
The metric in this parametrization becomes 
\begin{equation}\label{althopfspace}
  ds^2 = -\sinh^2\rho\, d\tau^2+d\rho^2+(d\psi+\cosh\rho\, d\tau)^2.
\end{equation}

For the timelike Hopf fibration we take the parametrization
\begin{equation}
g = \e^{\tau\hat\tau^2}\e^{\rho\hat\tau^1}\e^{\psi\hat\tau^2}.
\end{equation}
This leads to left-invariant one-forms $\omega^a$ given by
\begin{equation}
  \left\{
  \begin{array}{r@{\;}c@{\;}l}
    \omega^1 &=& \cos\psi\, d\rho + \sin\psi\sinh\rho\, d\tau\, , \\
    \omega^2 &=& d\psi + \cosh\rho\, d\tau\, , \\
    \omega^3 &=& -\sin\psi\, d\rho + \cos\psi\sinh\rho\, d\tau\, .
  \end{array}
  \right.
\end{equation}
The metric in this parametrization becomes 
\begin{equation}\label{althopftime}
  ds^2 = \sinh^2\rho\, d\tau^2+d\rho^2-(d\psi+\cosh\rho\, d\tau)^2\, .
\end{equation}

\section{$AdS_3$ Metrics in Poincar\'e Coordinates}
\label{app:poincare}

The metric for $AdS_3$ on a Poincar\'e patch can be written in the
form
\begin{equation}\label{poincare1}
  ds^2_{AdS_3} = \frac{dr^2}{r^2} + 2r\,dt\, du + \alpha\; du^2\, . 
\end{equation}
This is a valid $AdS_3$ metric for any value of $\alpha$, in
particular, after a rescaling, it can be chosen as $\alpha=\pm 1$ or
$\alpha=0$. We will show below that these three cases correspond to
the three different Hopf fibrations of $AdS_3$, which along the way
establishes them as metrics on $AdS_3$. For $\alpha=0$, the relation
with the lightlike Hopf fibration~\eqref{hopfnull} is obvious.  For
$\alpha=\pm1$ we can write this metric in the form
\begin{equation}\label{poincare2}
  ds^2 = \frac{dr^2}{r^2} - \alpha r^2dt^2 + \alpha(du+rdt)^2\, ,
\end{equation}
where we rescaled $t\to\alpha t$ for convenience.

\subsection{Coordinate Transformations in the Spacelike Case}

First consider the Poincar\'e metric~\eqref{poincare2} for $\alpha=1$.
Then consider the coordinate transformation
\begin{equation}\label{trafo1}
  \left\{
  \begin{array}{r@{\;}c@{\;}l}
    r  &=& \sinh\rho + \cosh\rho \cos\tau\, , \\
    rt &=& \cosh\rho \sin\tau\, , \\
    u  &=&\displaystyle \psi - 2\arctanh \Bigl(\e^{-\rho}\tan\frac\tau2\Bigr)\, . 
  \end{array}
  \right.
\end{equation}
From these transformations we derive 
\begin{equation}
  \frac{dr^2}{r^2} - r^2dt^2 = d\rho^2 - \cosh^2\rho\; d\tau^2\, ,\qquad
  du + rdt = d\psi + \sinh\rho\; d\tau\, . 
\end{equation}
Using this, we find that the metric~\eqref{poincare2} for $\alpha=1$
becomes the metric~\eqref{hopfspace} in global coordinates.  Because
we have established the separate identities above, we can also apply
this coordinate transformation to the five dimensional metric for
$j<1$.

\subsection{Coordinate Transformations in the Timelike Case}

Next we consider the Poincar\'e metric~\eqref{poincare2} for
$\alpha=-1$.  We take the coordinate transformation
\begin{equation}\label{trafo2}
  \left\{
  \begin{array}{r@{\;}c@{\;}l}
    r  &=& \sinh\rho + \cosh\rho \cosh\tau\, , \\
    rt &=& \cosh\rho \sinh\tau\, , \\
    u  &=&\displaystyle \psi - 2\arctan \Bigl(\e^{-\rho}\tanh\frac\tau2\Bigr)\, . 
  \end{array}
  \right.
\end{equation}
From these transformations we derive 
\begin{equation}
  \frac{dr^2}{r^2} + r^2dt^2 = d\rho^2 + \cosh^2\rho\; d\tau^2\, ,\qquad
  du + rdt = d\psi + \sinh\rho\; d\tau\, . 
\end{equation}
Using these relations we find that~\eqref{poincare2} for $\alpha=-1$
becomes the metric~\eqref{hopftime} in global coordinates.  Again
these coordinate transformations can be applied to the five
dimensional metric for $j>1$.

\subsection{Alternative Coordinate Transformations in the Spacelike Case}

We can also relate the Poincar\'e metrics to the alternative Hopf
fibrations.  First consider the Poincar\'e metric~\eqref{poincare2}
for $\alpha=1$.  The coordinate transformation we consider is
\begin{equation}\label{trafo4}
  \left\{
  \begin{array}{r@{\;}c@{\;}l}
    r  &=& \cosh\rho + \sinh\rho \cosh\tau\, , \\
    rt &=& \sinh\rho \sinh\tau\, , \\
    u  &=&\displaystyle \psi + 2\arctanh \Bigl(\e^{-\rho}\tanh\frac\tau2\Bigr)\, .  
  \end{array}
  \right.
\end{equation}
From these transformations we derive 
\begin{equation}
  \frac{dr^2}{r^2} - r^2dt^2 = d\rho^2 - \sinh^2\rho\; d\tau^2\, ,\qquad
  du + rdt = d\psi + \cosh\rho\; d\tau\, . 
\end{equation}
Hence the metric~\eqref{poincare2} for $\alpha=1$ becomes the
metric~\eqref{althopfspace} in global coordinates.

\subsection{Alternative Coordinate Transformations in the Timelike Case}

Next consider the Poincar\'e metric~\eqref{poincare2} with
$\alpha=-1$.  We consider the coordinate transformation
\begin{equation}\label{trafo3}
  \left\{
  \begin{array}{r@{\;}c@{\;}l}
    r  &=& \cosh\rho + \sinh\rho \cos\tau\, , \\
    rt &=& \sinh\rho \sin\tau\, , \\
    u  &=&\displaystyle \psi + 2\arctan \Bigl(\e^{-\rho}\tan\frac\tau2\Bigr)\, . 
  \end{array}
  \right.
\end{equation}
From these transformations we derive 
\begin{equation}
  \frac{dr^2}{r^2} + r^2dt^2 = d\rho^2 + \sinh^2\rho\; d\tau^2\, ,\qquad
  du + rdt = d\psi + \cosh\rho\; d\tau\, . 
\end{equation}
Hence~\eqref{poincare2} with $\alpha=-1$ becomes the 
metric~\eqref{althopftime} in global coordinates.

\end{document}